\documentclass[useAMS,usenatbib]{mn2e}
\usepackage{graphicx}
\usepackage{longtable}
\usepackage{subfigure}
\usepackage{amsmath}
\usepackage{epsfig}
\usepackage{supertabular}
\usepackage{epstopdf}
\usepackage{amssymb}

\title[On the surface density of dark matter haloes]{ On the surface density of dark matter haloes }

\author[A. Saburova, A. Del Popolo]{A. Saburova$^{1}$\thanks{Corresponding author\,: {\tt saburovaann@gmail.com}}, A. Del Popolo$^{2,3}$\\
$^1$ Moscow M.V. Lomonosov State University, Sternberg Astronomical Institute, Universitetskij pr., 13, 119992,  Moscow, Russia \\
$^2$ Dipartimento di Fisica e Astronomia, Universita di Catania, and INFN Sezione Catania, Via S. Sofia 64, 95123 Catania, Italy\\
$^3$ International Institute of Physics, Universidade Federal do Rio Grande do Norte, 59012-970 Natal, Brazil
}

\begin{document}

\pagerange{\pageref{firstpage}--\pageref{lastpage}} \pubyear{2002}

\maketitle

\label{firstpage}

\begin{abstract}

In this article, we test the conclusion of Donato et al. (2009) concerning the universality of the DM halo surface density $\mu_{0D}=\rho_0r_0$. 
According to our study, the dispersion of values of $\mu_{0D}$ is twice higher than that found by Donato et al. (2009). We conclude, in contrast 
with Donato et al. (2009), that the DM surface density and its Newtonian acceleration are not constant but correlate with the luminosity, morphological type, $(B-V)_0$ colour index, and the content of neutral hydrogen. These DM parameters are higher for more luminous systems of early types with red colour and low gas content. We also found that the correlation of DM parameters with colour index appears to be the manifestation of a stronger relation between DM halo mass and the colour of a galaxy.  This finding is in agreement with cosmological simulations (Guo et al, 2011). 
These results leave little room for the recently claimed universality of DM column density. We also found that isolated galaxies in our sample (contained in the Analysis of the interstellar Medium of
Isolated GAlaxies (AMIGA) catalogue) do not differ significantly in their value of $\mu_{0D}$ from the entire sample. Thus, since the AMIGA catalogue gives a sample of galaxies that have not interacted with a significant mass neighbour in the past 3 Gyr, the difference between the
systems with low and high values of $\mu_{0D}$ is not related to the merging events during this period of time.
\end{abstract}


\section{Introduction}

Although the $\Lambda$CDM model has been shown in the last decade
to predict with high accuracy observations on cosmological \footnote{
However, even at these scales the $\Lambda$CDM paradigm is affected by the cosmological constant problem \citep{Weinberg, Astashenok}, and the ``cosmic coincidence problem''and the total ignorance on the origin of the accelerated expansion of the Universe (\citealt{DelPopoloPace2013, DelPopoloPaceLima2013}, \citealt{DelPopoloPaceMaydanyuk2013}).}, and intermediate scales (\citealt{Spergel}; \citealt{Kowalski}; \citealt{Percival}; \citealt{Komatsu}; \citealt{DelPopolo2007ARep}; \citealt{2013AIPC15482D}; \citealt{2014IJMPD2330005D}), the quoted paradigm is not successful to the same extent in
explaining observations on smaller scales, from tens of pc to some
kpc. A well-known problem of the $\Lambda$CDM model on small scales
is the discrepancy in the predicted cuspy density profiles of galaxies
in N-body simulations (\citealt{nfw1996}; \citealt{nfw1997}; \citealt{Navarro2010}), and the observed cored profiles in dwarf spirals,
dwarf spheroidals (dSphs) and low-surface-brightness galaxies (\citealt{Burkert1995}; \citealt{deBloketal2003}; \citealt{Swatersetal2003}; \citealt{DelPopolo2009} (DP09), \citealt{DelPopoloKroupa2009}; \citealt{DelPopolo2010MNRAS}, \citealt{2012MNRAS419971D} \citealt{2012MNRAS42438D} (DP12a, DP12b); \citealt{DelPopoloHiotelis2014}; \citealt{2011AJ14224O}; \citealt{2011AJ141193O}; \citealt{KuziodeNaray2011}). This problem, known as the cusp/core problem (\citealt{Moore1994}; \citealt{Flores1994}) is flanked by the ``missing satellite problem" dealing
with the discrepancy between the number of subhaloes predicted in
N-body simulations (\citealt{Klypin1999}; \citealt{Moore1999}) and that
observed. The subhaloes predicted in simulations are not only more
numerous than satellites of galaxies like the Milky Way (MW),
but also more dense than expected (\citealt{2011MNRAS415L40B}; \citealt{2012MNRAS4221203B}). Smooth Particle Hydrodynamics (SPH) simulations of a galaxy with not enough large resolution, and not taking properly account of feedback mechanisms, like, for example, supernovae explosions, produce discs too small with respect to the discs of spiral galaxies (\citealt{2001MNRAS3261205V}; \citealt{2008ASL17M}). The angular momentum distributions
produced are not in agreement with those observed, since
angular momentum is lost during repeated collisions through dynamical
friction. Additional concerns come from the flatness at the
low-mass end of the luminosity function (e.g., \citealt{2009MNRAS3991106M}), the galactic stellar mass function (\citealt{2008MNRAS388945B}; \citealt{Li2009}), and 
the HI mass function (\citealt{Martin2010}, as well as the ``void phenomenon" (\citealt{Peebles2001}; \citealt{Tinker2009}), and the discrepancy in the sizes of mini-voids obtained in $\Lambda$CDM simulations and those observed (\citealt{Tikhonov2009}).

%Even if 
Several solutions have been proposed to the previous problems,
based on modifying the particles constituting dark matter (DM) (\citealt{2000ApJ542622C}; \citealt{2001ApJ551608S}; \citealt{2000NewA5103G}; \citealt{2000ApJ534L127P}), modifying the power spectrum (e.g. \citealt{2003ApJ59849Z}), modifying the theory of gravity (\citealt{1970MNRAS1501B}; \citealt{1980PhLB9199S}; \citealt{1983ApJ270365M}; \citealt{1983ApJ270371M}; \citealt{Ferraro2012}), 
%they are somehow not welcome since they imply to reject a successful paradigm like the $\Lambda$CDM model for a new paradigm less well studied than it.
or astrophysical solutions based on mechanisms that ``heat" DM\footnote{ The two main mechanisms are a) supernovae feedback  (\citealt{1996MNRAS283L72N}; \citealt{Gelato1999}; \citealt{Read2005}; \citealt{Mashchenko2006}, \citealt{Mashchenko2008}; \citealt{Governato2010}; \citealt{Zolotov2012}), and b) transfer of energy and angular momentum from baryons to dark matter through dynamical friction (El-Zant et al. 2001, 2004; \citealt{2008ApJ685L105R}, \citealt{2009ApJ7021250R}; \citealt{DelPopolo2009}; \citealt{Cole2011}; \citealt{DelPopoloHiotelis2014}; \citealt{2014JCAP04021D}) }.

Delegating the solutions of the previous problems to the poorly
understood and complex phenomena happening in dense plasma
and to baryonic physics in general is more natural, but not easy.

Often in astrophysics, scaling relations are of noteworthy help in
understanding complex phenomena and in constraining DM properties
and formation scenarios. 

In this context, \cite{Kormendy2004} found several interesting
relations among DM halo parameters, obtained through mass
modelling of the rotation curves of 55 galaxies using a pseudo-isothermal
profile as a fitting profile. Of particular interest is the
quantity $\mu_{0D} =\rho_{ 0~p.i.} r_{0~p.i.}$, where $r_{0~p.i.}$ is the core radius of the pseudo-isothermal profile, and $\rho_{0~p.i.}$ its central density. They found that the
previous quantity, the surface density of DM haloes, is independent
of galaxy luminosity in the case of late-type galaxies and has a value
of $\simeq 100 M_{\odot} pc^{-2}$. The previous result was extended by \cite{Donato} (hereafter D09) using $\simeq  1000$ spiral galaxies, and dwarfs rotation curves and the weak lensing of spirals and ellipticals; they
claimed a quasi-universality of the central surface density of DM
haloes. 
%of $\simeq  1000$ spiral galaxies, the mass models of individual dwarf and spiral galaxies and the weak lensing signal of elliptical and spirals,

Prompted by the previous claim, \cite{Milgrom2009} showed that modified Newtonian dynamics (MOND) predicts, in the Newtonian
regime, a quasi-universal value of $\mu_{0D}$ for every different kind of
internal structure and for all masses. The quoted quasi-universality
was extended by \cite{Gentile2009} (G09) to  luminous matter surface density. Namely, they claimed that the ratio of luminous
to dark matter is constant within one halo scalelength. 

The D09 and G09 results are based on the assumption that all
galaxies, from dwarfs to ellipticals, are fitted by a Burkert DM halo
density profile: \begin{equation}\label{Burkert}
\rho(r)=\frac{\rho_0 r_0^3}{(r+r_0)(r^2+r_0^2)}
\end{equation}

The Burkert profile usually gives a good fit to the rotation curves
of dwarfs and LSBs (\citealt{Gentile2004}, \citealt{Gentile2007}; Del Popolo 2009), but exceptions exist (\citealt{Simon2005}; \citealt{THINGS}; \citealt{2012MNRAS42438D}). For example \cite{Simon2005} showed that in the case of 
NGC 2976, 4605, 5949, 5693 and 6689, the inner slope of the density profile spans a range $ \simeq 0$ (NGC 2976) to $-1.28$ (NGC 5963). \cite{THINGS} found that brighter, larger galaxies with $M_{\rm B}< -19$ have density profiles well fitted by both cuspy profiles
and cored ones, while less massive galaxies with $M_{\rm B}> -19$ are best fitted by cored profiles. Even the MW dSphs could have cuspy profiles. According to \cite{Strigari2010}, and \cite{Breddels2013}, Fornax has a cuspy profile, while it is cored according to other authors (\citealt{Jardel2012}; \citealt{Jardel2013};
\citealt{JardelGebhardt2013}; \citealt{2013MNRAS4333173B}; \citealt{Walker2011}; \citealt{Amorisco2012}; \citealt{Battaglia2008}; \citealt{Agnello2012}). 

Concerning elliptical galaxies, the analysis is more complex than
in the case of spirals. Several methods are used to study the content
and distribution of DM, for example the virial theorem (see \citealt{2014IJMPD2330005D}), the X-ray properties of the emitting hot gas (see \citealt{Buote2012} for a review, \citealt{Nagino2009}), the
dispersion velocities of kinematical tracers and stellar dynamics (see \citealt{Gerhard2010} for a review; \citealt{Napolitano2011}) and combining weak and strong lensing data (see \citealt{Ellis2010} and \citealt{Treu2010} for a review).

Nowadays we know from X-ray, lensing observations and stellar
dynamics studies that elliptical galaxies are surrounded by large
DM haloes, even if studies of some years ago, based on Jeans analysis,
concluded that the quantity of DM in ellipticals was meagre (\citealt{Romanowsky2003})\footnote{Romanowky's conclusion that ellipticals contained a dearth of DM, based on a falling velocity dispersion, was later revised by \cite{Dekel2005} through simulations of merging spiral galaxies, which again showed a falling velocity dispersion, even if the galaxies contained plenty of DM.}

In the present day, Chandra, XMM, and Suzaku have measured
the temperature profiles of many galaxies accurately, leading to precise
constraints on the $M/L$ ratios of those galaxies (see \citealt{Buote2012} and references therein). The increase of $M/L$ with radius,  the dynamics of satellite galaxies (\citealt{Prada2003}) and weak lensing (\citealt{Hoekstra2005}; \citealt{Kleinheinrich2006}; \citealt{Mandelbaum2006}; \citealt{Heymans2006}) show that the DM
halo mass is much larger than the stellar mass. However, doubts
remain as to the functional form of the best-fitting DM profile (e.g.
Navarro-Frenk-White (NFW), consistent with X-ray data, and the Sersic-Einasto profile (\citealt{Merritt2006}; \citealt{Graham2013}) consistent with dissipationless simulations).

The previous discussion puts obvious doubts on the conclusions of D09 and G09, based on the assumption that all the galaxies in
their sample are well fitted by cored profiles, namely the Burkert
profile.

Several authors have studied the surface density of galaxies,
reaching opposite results to those of D09 and G09, namely that the surface density is not universal. 
\cite{Napolitano2010} showed that the projected density of local early-type
galaxies within the effective radius is larger than that of dwarfs
and spirals. This systematic increase with the mass of the halo was
also noticed by \cite{Boyarsky}. The \cite{Boyarsky} sample was larger than those of D09 and G09, including groups and clusters. The dark matter column density, $S$, defined by \cite{Boyarsky} as 
\begin{equation}    
\label{eq:Sbar}
S = \frac2{r_\star^2} \int^{r_\star}_0 rdr
\int  dz \rho_{DM}(\sqrt{r^2+z^2})
%\label{eq:column}
\end{equation}
is proportional to the mean surface density in $r_\star$ (i.e., $\rho_{\ast} r_{\ast}$ is equal to $\rho_0 r_0$ for the Burkert density profile).  Here $r_\star$ is the characteristic
scale at which the inner slope of the DM density profile changes
towards its outer asymptotic value, with an average density $\rho_{\ast}$ within this radius. The dark matter column density, $S$, is given in \cite{Boyarsky} by
\begin{equation}
\log S= 0.21 \log \frac{M_{halo}}{10^{10} M_{\odot}}+1.79
\end{equation}
with S in $M_{\odot}pc^{-2}$. 
\cite{CardoneTortora2010}, 
based on central velocity
dispersion of local galaxies and strong lensing at intermediate
redshift, have shown that the Newtonian acceleration and column
density correlate with different quantities: the visual luminosity $L_V$, the effective radius $r_{e}$, the stellar mass $M_{\ast}$ and the halo mass $M_{200}$ in agreement with \cite{Boyarsky} and in disagreement with the results of D09 and G09. 
%Cardone \& Tortora (2010) (CT10), by modeling the dark halo with a Navarro - Frenk -White profile and assuming a Salpeter initial mass
%function (IMF) to estimate stellar masses, found that the column density and the Newtonian acceleration within the halo characteristic radius $r_s$ and %effective radius $R_{eff}$ are not universal quantities, but correlates with the luminosity $L_V$ , the stellar mass $M_{\ast}$ and the halo mass %$M_{200}$, contrarily to what is expected from D09 and G09, and in agreement with B09.

\cite{Napolitano2010} calculated $<\rho_{\rm DM}> r_{\rm e}$, which is proportional to the column density $S$, for a large sample of elliptical galaxies. They found that early-type galaxies violate the constant-density
scenario. 

\cite{2013MNRAS4291080D} found a correlation of the surface density with $M_{200}$, in agreement with \cite{Napolitano2010}, \cite{CardoneTortora2010}, and \cite{Boyarsky}, but
with a smaller scatter, and they also found a correlation between
baryon column density and mass.

The results of \cite{CardoneDelPopolo2012} followed as close as possible the analysis of D09 end G09, but improving the way the
halo models were estimated\footnote{D09 and G09 evaluated the haloes parameters adjusting previous results published in literature, while \cite{CardoneDelPopolo2012} estimated the halo model parameters by fitting the rotation curves.} and doubling the D09, G09 sample, allowing for an investigation of selection effects.  
%Third, we adopt a Bayesian fitting procedure to infer a realistic estimate of the errors on the quantities of interest also taking into account the %uncertainties on the mass - to - light ratio
They found a
correlation between the Newtonian acceleration and the virial mass $M_{\rm vir}$, testifying against the universality claimed in the D09 and G09 articles.
%Whether the quoted tensions are an evidence for a failure of the $\Lambda$CDM scenario or for a wrong modeling of the baryons physics in 
%the simulations is still a debated issue. As a valuable help, one can investigate empirical correlations among dark matter and stellar properties as a %way to constrain the galaxy formation scenarios.
%
%, whose validity has been much debated (Bosma 2004, Palunas and
%Williams 2000). 
%
%This was interpreted as a close correlation between the enclosed surface densities of luminous and dark matter in galaxies and a similar 
%value to those obtained by D09.
%Donato et al. (2009), Gentile et al. (2010).
%recent claims in literature.

In the present article, we extend the sample used in \cite{CardoneDelPopolo2012} to study whether new data change the \cite{CardoneDelPopolo2012} and \cite{2013MNRAS4291080D}  conclusions. Moreover, we expand the previous analysis, studying whether the
surface density and Newtonian acceleration correlate with luminosity,
morphological type, colour index and neutral hydrogen content.

%In order to try to discriminate among these results and find an explanation and analytical derivation of the surface density of haloes, we analyze the %problem using the secondary infall model (SIM) introduced in Del Popolo (2009) (hereafter DP09), taking into account ordered and random angular %momentum, dynamical friction, and baryon adiabatic contraction. 

The article is organized as follows. In Section 2, we discuss the
methods utilized to obtain the parameters of DM haloes in the current
article. In Section 3, we discuss the uncertainties in the DM parameter
estimates, including the possible effect of non-universality
of the initial mass function (IMF); results are given in Section 4,
while Section 5 is devoted to a discussion and conclusions.

\section{On the methods of  DM halo parameters estimation}

As previously reported, it was claimed by D09 that the surface
density of DM haloes $\mu_{0D} =\rho_0 r_0$ is constant for galaxies of different
types and luminosities. The universality of the surface density
of DM haloes is intriguing, because according to D09 it persists
for dark halo parameters obtained using different methods. In fact,
different approaches could give significantly different results even
when applied to the same systems. For disc galaxies, one can perform
the decomposition of the rotation curve into the contributions
of bulge, gaseous and stellar disc and DM halo: \begin{equation}    
v^2(r)=v^2_{bulge}(r)+v^2_{disc}(r)+v^2_{DM~halo}(r).
\end{equation}  
\subsection{Best-fitting model of the rotation curve}
Using the so-called best-fitting model, based on minimizing the deviation
of the model rotation curve from the observed one, could
give rise to different parameters of the dark matter halo for the same
object. This problem can have several solutions, due to the uncertainty
of the baryonic contribution to the total mass of a galaxy. To
avoid this degeneracy, one should introduce additional information
when performing mass modelling of disc galaxies.  To do this, one
can fix the surface density of the baryonic matter and consequently the contributions to the rotation curve of the bulge $v^2_{bulge}(r)$ and the disc: \begin{equation}    
v^2_{disc}(r)= K_d^2 y^2 (I_0(y)K_0(y)-I_1(y)K_1(y)), \end{equation}     where
$K_d=\sqrt{4\pi G \sigma_{0d} r_d}$ , $y=\frac{r}{2r_d}$,  $K_n$ and $I_n$ are modified Bessel functions and $\sigma_{0d}$ and $r_d$ are the central surface density and exponential
scalelength of an exponential disc. The bulge contribution $v^2_{bulge}(r)$ is modelled in the same way as for the disc: when the density
distribution is known (it is usually accepted to be proportional
to the surface brightness, e.g. King's or Sersic's law), one can obtain the gravitational potential $\phi$ using Poisson's equation: \begin{equation}    
\Delta \phi= -4 \pi G \rho. \end{equation}   After that, one can calculate the contribution to the circular
velocity: \begin{equation} v^2/r=\partial \phi/\partial r. \end{equation} 
When the contributions of bulge and disc to the rotation curve are
known, the model of the rotation curve is constructed by choosing
the parameters of the DM halo minimizing the remaining difference
between the baryonic and observed rotation curves. To fix the baryonic
density, one uses additional observational data. In the absence
of additional information on the baryonic density, one can use the
maximum disc model, in which one adopts the highest possible
contribution of baryonic matter to the rotation curve. 
 \subsection{Photometric approach}
To obtain the surface density of the stellar disc and bulge, one
can use stellar population synthesis models together with surface
photometry (see e.g. \citealt{THINGS}). In this approach, the stellar
surface density radial profile is calculated from the profile of surface
brightness multiplied by the mass-to-light ratio, which is obtained
from the observed colour index and stellar population synthesis
model $M/L$ -- colour relation (see e.g. \citealt{bdj}): \begin{equation} \sigma_{0d}=M/L \cdot I_{0d}.\end{equation} The density of the gaseous disc could be found from radio observations.

One can also perform a more detailed analysis to put constraints
on the stellar density distribution, as was done in \cite{Kasparova2014}, where the density of the stellar bulge and disc were determined
from modelling of long-slit stellar spectra, together with the
spectral energy distribution.

\subsection{Method based on spiral-arm density-wave theory}
 Another way to fix the baryonic density in mass decomposition
is to take into account the presence of spiral structure (see e.g. \citealt{Athanassoula}, \citealt{Fuchs2003}). Density wave
theory allows us to make a prediction of the number of spiral
arms. The circumferential wavelength of the spiral density waves is
connected to the critical wavelength, which in turn is related to the
density of the disc: \begin{equation}    \lambda_{crit}=4\pi^2 G \sigma_d/\kappa^2, \end{equation}     where $\sigma_d$ is the surface density of the disc, $G$ is gravitational constant and $\kappa=2\Omega\cdot\sqrt{(1+(r/2\Omega)\cdot(d\Omega/dr))}$ is 
the epicyclical frequency, determined by the angular velocity $\Omega$. The number of spiral
arms $m$ is connected to the critical wavelength through the equation $m=2\pi r/X \lambda_{crit}$, where $X$ is the coefficient depending on the slope of the rotation curve. Thus knowing the number of spiral arms allows
one to determine the density of the disc, which can be fixed
in rotation curve decomposition. This in turn allows one to narrow
the range of possible parameters of the DM halo.
\subsection{The approach based on the disc marginal gravitational
stability}
In another approach, one can use the marginal gravitational stability
criterion (see \citealt{Bottema}; \citealt{Zasov2011}; \citealt{Saburova2011}; \citealt{Saburova2012}) to obtain the density of the disc. For a thin disc with parameters
slowly varying with $r$, the local critical value of the radial velocity
dispersion of stars is defined by the Toomre criterion: \begin{equation}    c_r=Q_Tc_T, \end{equation}     where $c_T ={3.36 G \sigma_d}/{\kappa}$ and parameter $Q_T$ depends on the radius and lies in the range $1<Q_T<3$ (see \citealt{Khoperskov2003}). Thus, if the radial stellar velocity dispersion $c_r$ is known, one
can assume that the disc is in a marginally stable state, estimate the
surface density of the disc independently and then use it in mass
modelling.
\subsection{Techniques applied to ellipticals and dwarf spheroidals}

To determine the DM halo parameters of elliptical galaxies, one can
use the X-ray properties of hot gas (see e.g. \citealt{Humphrey2006}), planetary nebulae or stellar kinematical data (see e.g. \citealt{2009MNRAS393329N} and \citealt{2009MNRAS398561W}). The DM properties of dwarf
spheroidal galaxies could also be obtained from kinematical data
by applying Jeans analysis to the velocity dispersion profiles (see \citealt {2009PhDT124F}). Wu (2007) performed modelling of dSph
galaxies as axisymmetric stellar systems in spherical potentials,
which are in dynamical equilibrium without significant external
tidal forces.
\section{On the uncertainty of the estimate of DM parameters}
As different approaches use different assumptions, they can give
different results. A good example of this could be LSBs. Based on
stellar population models with standard Salpeter-like stellar IMF,
LSBs are believed to be dark-matter-dominated systems. However,
some independent estimations of their disc masses give evidence
that LSBs may have extremely high disc mass-to-light ratios in the
presence of a dark halo of moderate mass (see e.g. \citealt{Saburova2011}, \citealt{Fuchs2003}), which can be explained if the IMF of LSBs
is not universal but rather bottom-heavy (with an excess of low mass
stars, which contribute to the mass of the galaxy but have
no significant contribution to the luminosity: see \citealt{Lee2004}). Different assumptions made for these objects lead to a different
understanding of their nature and consequently to different DM
parameters.

\begin{figure*}
\includegraphics[width=11cm,keepaspectratio]{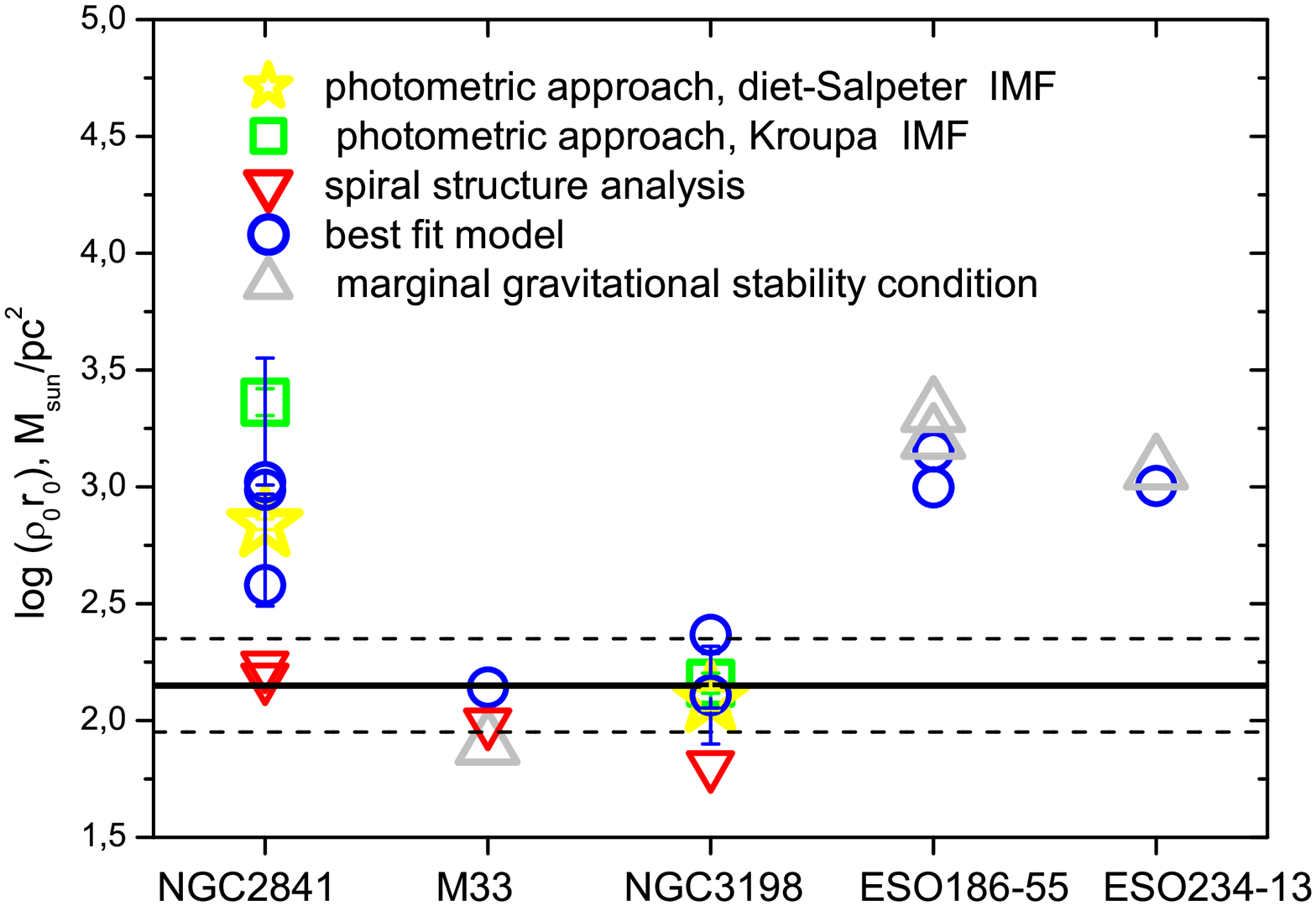}
\caption{A comparison of DM halo surface densities $\log \mu_{0D}$ obtained using different approaches for the galaxies. Solid and dashed black lines represent the value of $\log \mu_{0D} =2.15\pm 0.2$ obtained by 
D09.}
\label{comparison}
\end{figure*}
\begin{table*}
\begin{center}
\caption{ DM halo parameters obtained using different approaches. \label{comparison_table}}
    \begin{tabular}{lllll}
    \hline
Galaxy&Aproach&$\rho_0$&$r_0$&Source\\
&&$M_{\odot}/pc^3$&kpc&\\
\hline
 \hline
(1)&(2)&(3)&(4)&(5)\\
 \hline
NGC2841& spiral structure analysis&0.005&30&\cite{Athanassoula}\\
& spiral structure analysis&0.004&32.6&\cite{Athanassoula}\\
& photometric approach, Kroupa IMF&3.215$\pm0.372$&0.63$\pm0.04$&\cite{THINGS}\\
& photometric approach, Salpeter IMF&0.299$\pm0.014$&2.03$\pm0.05$&\cite{THINGS}\\
& best fit&0.675$\pm0.75$&1.36$\pm0.75$&\cite{THINGS}\\
& best fit&0.09 &3.7&\cite{Kasparova2012}\\
& best fit&0.16$^{+0.01} _{-0.01}$&5.91$^{+0.17}_{-0.08}$&\cite{CardoneDelPopolo2012}\\

\hline
M33& spiral structure analysis&0.011&8&\cite{Athanassoula}\\
& best fit&0.011&13&\cite{Corbelli}\\
& gravitational stability condition&0.009&7.5&\cite{Saburova2012}\\
\hline
NGC3198
& spiral structure analysis&0.002&23.4&\cite{Athanassoula}\\
& photometric approach, Kroupa IMF&0.047$\pm0.004$&2.72$\pm0.13$&\cite{THINGS}\\
& photometric approach, Salpeter IMF&0.033$\pm0.003$&3.22$\pm0.16$&\cite{THINGS}\\
& best fit&0.047$\pm0.011$&2.71$\pm0.33$&\cite{THINGS}\\
& best fit&0.064 &3.19&\cite{Kasparova2012}\\
& best fit&0.013$^{+0.006} _{-0.005}$&9.81$^{+1.55}_{-1.71}$&\cite{CardoneDelPopolo2012}\\

\hline
ESO186-55&best fit&4.3&0.33&\cite{Shchelkanova}\\
& best fit&1.40&0.71&\cite{Shchelkanova}\\
& gravitational stability condition &5.90&0.27&\cite{Shchelkanova}\\
& gravitational stability condition &16.0&0.13&\cite{Shchelkanova}\\
\hline
ESO234-13&best fit&2.1&0.48&\cite{Shchelkanova}\\
& gravitational stability condition &3.10&0.38&\cite{Shchelkanova}\\

\hline

 \end{tabular}
 
 \end{center}
\end{table*}

To explore how uncertain the parameters of a DM halo might
be, we compared the DM surface densities obtained using different
approaches for three well-studied high-surface-brightness galaxies
-- NGC 2841, M33 and NGC 3198 -- and two LSBs -- ESO 186–55
and ESO 234–13 (see Fig. \ref{comparison}). Different colours in Fig. \ref{comparison} correspond
to different methods of rotation curve decomposition, which include
best-fit modelling, an approach based on spiral structure analysis, a
method based on the marginal gravitational stability condition and
the photometric approach. For the source of the DM parameters, see Table \ref{comparison_table}. It follows from Fig. \ref{comparison} that the dispersion of DM surface
density could reach $\approx$ 1.2 dex for the same galaxy (for NGC 2841).
This is mostly due to the fact that every method has its own uncertainty.
For example, the method based on the marginal gravitational
stability condition gives only the upper limit of the surface density
of the disc, because the disc could have a stellar velocity dispersion
higher than is needed for the gravitationally stable state and hence
this gives the lower limit of the DM halo density.

In best-fit modelling, due to the unknown contribution of baryonic
matter to the total mass of the galaxy, one can obtain several
solutions of the problem of rotation curve decomposition that reproduce
the observed rotation curve equally well. 
\subsection{On the non-universality of the IMF and its possible effects
on DM halo parameter estimates}
The photometric approach is based on the assumption that the IMF
is universal. However, there is evidence that the IMF could change
in early-type galaxies with central stellar velocity dispersion, mass (see, e.g. \citealt{LaBarbera}, \citealt{Ferreras2013}) and stellar mass-to-light ratio (\citealt{Cappellari2012}). \cite{Tortora2013} concluded that early-type galaxies with high
central stellar velocity dispersion tend to have an excess of low mass
stars relative to spirals and early-type galaxies with low velocity
dispersions. Recently, \cite{2014arXiv14046533M} found
that the IMF could also vary with radius. Systematic changes of
the IMF could lead to systematic changes of stellar mass estimations
and consequently to a difference in DM surface densities. The
DM halo surface densities calculated from the DM halo parameters
obtained by \cite{THINGS}, taking into account transformations
to the Burkert DM density distribution (see Section 4
for the equations), are systematically different for Kroupa (\citealt{imf2}) and diet-Salpeter (\citealt{imf}) IMFs. The mean value of the pseudo-isothermal DM halo density for the diet-Salpeter
IMF is $log(\mu_{0D}) =1.94\pm0.35$, whereas for Kroupa IMF: $log(\mu_{0D}) =2.19\pm0.47$. The same conclusion is valid for a cuspy
NFW profile: the DM surface density is higher for models with
Kroupa IMF, $log(\mu_{0D}) =2.32\pm0.47$, than for the Salpeter IMF, $log(\mu_{0D}) =2.11\pm0.44$.  The surface density of the DM halo is
higher for the Kroupa IMF, because it corresponds to 1.4 times
lower stellar mass. One should mention, however, that the difference
between the DM surface densities obtained for the different
IMFs is quite moderate and the estimates are consistent within the
errors. 

Another possible effect of IMF variation could appear, due to the
fact that mass models with lower stellar $M/L$s are often obtained for the cuspy profiles due to the steeper mass distribution (see \citealt{THINGS}). Lower stellar $M/L$ could be a result of a lower amount
of low mass stars and correspondingly different IMF. To evaluate
the influence of this effect on DM surface densities, we calculated
the values of $log(\mu_{0D})$ from the DM parameters of the best-fitting
models of \cite{THINGS} for NFW and pseudo-isothermal density distributions. In the calculation, we used transformations to
the Burkert profile given in Section 4. We obtained the following
values of the mean surface densities: $log(\mu_{0D}) =2.19\pm0.41$ for the pseudo-isothermal density distribution and $log(\mu_{0D}) =2.28\pm0.43$ for the NFW profile. The DM surface density is slightly higher for
the NFW case in comparison with the pseudo-isothermal distribution.
However, the difference is within the errors.

In the current article, for the DM parameters obtained by the
photometric approach, the IMF is universal and has a diet-Salpeter
form (\citealt{imf}). According to \cite{Tortora2013}, one can
expect that there could be a systematic change of the IMF from early
to late-type galaxies. Early-type galaxies with high central velocity
dispersion could possess higher stellar $M/L$ in comparison with
spirals. Having in mind that we use the diet-Salpeter IMF, which
corresponds to models that are close to the maximum disc models (see \citealt{THINGS}), we can expect that we will overpredict the
stellar density for late-type galaxies. At the same time, we obtain a
correct or even underestimated stellar surface density for early-type
galaxies. Bearing in mind the discussion given above, this could
mean the underestimation of DM surface density for spirals and
overprediction for early-type galaxies. This could be the origin of
the systematic effects in the relations discussed below. To avoid
systematics that could be a result of the possible non-universality of
the IMF, we used not only photometric but also other approaches for
galaxies in the same range of luminosities, types and colours. We
also calculated the averaged DM parameters for those galaxies for
which values of DM parameters were obtained in different articles by different approaches. As can be seen below, all our conclusions
are valid for galaxies with averaged estimates. This could mean that
the non-universality of the IMF has no significant influence on our
main conclusions.

\section{Results}
As follows from the previous section, different approaches could
lead to different DM parameters. As every method has its own weak
point, it is important to use DM parameters obtained by different
approaches in order to compensate for the uncertainties in each
particular approach. Use of different approaches allows us to make
general conclusions on the constancy of DM halo surface density.
Having that in mind, we tested the constancy of DM surface density
for the DM parameters available in the literature and obtained
using the methods described above. When several DM parameter
estimates were available for a galaxy, we averaged these parameters,
taking into account the error bars and the difference in the
adopted distances. When errors in DM parameters were absent in
the source article, we assumed that the error is 30 per cent during
averaging. The properties of the sample galaxies are given in Table \ref{sample}. We used a
sample of galaxies and some methods that differ from those utilized
in D09. Moreover, in contrast to D09, we used averaged DM
parameters for those galaxies with several estimates of the DM parameters
available in the literature, which allows us partly to avoid
the systematic effects of the methods. The sample consists of 211
galaxies of different types. We used the following approaches to the
DM parameter estimation described in Section 2:\\  (i) -- best fit modelling of the rotation curves (\citealt{Shchelkanova}, \citealt{Barnes}, \citealt{Kasparova2012}, \citealt{Chemin2006}, \citealt{vanEymeren2009}, \citealt{Weijmans2008}, \citealt{Spano2008}, \citealt{deBlokBosma2002}, \citealt{vandenBosch2001}, \citealt{Corbelli}, \citealt{CardoneDelPopolo2012}
);\\ (ii) -- maximum disc model of the rotation curve ( \citealt{Swatersetal2003},  \citealt{Swaters2011});\\
(iii) -- method based on the stellar population synthesis model
colour-mass-to-light ratio relation (\citealt{THINGS} (THINGS), \citealt{Lelli}\footnote{The parameters given by \cite{Lelli} are made by maximizing the disc contribution to the rotation curve, however the disc mass-to-light ratios are very close to the ones that could be found from the observed colour indices and  \cite{bdj} $M/L$-colour relations obtained for diet-Salpeter IMF (B-V=0.9 for Malin1 corresponds to $M_{HSB}/L_R=3.4$ (3.3 in \citealt{Lelli}) for NGC7589 $B-R=1.47$ (\citealt{Galaz2006}) corresponds to $M_{HSB}/L_R=3.2$ (2.6 in \citealt{Lelli}) ). }, \citealt{Weldrake2003}, \citealt{Weijmans2008}\footnote{We used two estimates of the DM parameters given by \cite{Weijmans2008}. The first one corresponds to the maximum disc model. In this model the mass-to-light ratio of stellar population is close to that found for Salpeter IMF. However the authors argue that in this case the mass-to-light ratio is too high. Thus we also used their second model in which the stellar mass-to-light ratio corresponds to the Kroupa IMF and averaged these two estimates.}, \citealt{Kuzio2008}, \citealt{Begum2004});\\ (iv) -- approach based on the spectral energy distribution and longslit
spectra modelling (\citealt{Kasparova2014});\\ (v) -- method relying on the marginal gravitational stability of the
disc (\citealt{Shchelkanova},  \citealt{Saburova2012});\\ (vi) -- approach taking into account the presence of spiral structure (\citealt{Athanassoula});\\ (vii) -- method in which the stellar mass-to-light ratio is obtained
from the colour index (the so-called ''Bottema disc'' of \citealt {deBlok1997});\\ (viii) --  method based on the X-ray properties of hot gas (\citealt{Humphrey2006});\\ (ix) -- Jeans analysis based on planetary nebulae or stellar kinematical
data (\citealt{2009MNRAS393329N}, \citealt{Napolitano2011}, \citealt{2009MNRAS398561W}, \citealt {2009PhDT124F}); \\ (x) -- modelling reproducing the observed velocity dispersion profiles
and number density profiles (\citealt{2007astroph2233W}). 

In all sources specified above, the authors used four types of DM
density profile. 

{\it (i) Pseudo-isothermal profile:}   \begin{equation}    \rho_{p.i.}(r)=\frac{\rho_{0~p.i.}}{(1+(r/r_{0~p.i.})^2)}.
\end{equation} This type of distribution was used in the following articles: \cite{Barnes}, \cite{Kasparova2012}, \cite{Kasparova2014},  \cite{vanEymeren2009}, \cite{Weijmans2008}, \cite{deBlokBosma2002}, \cite{Saburova2012}, \cite {deBlok1997}, \cite{Begum2004}, \cite{Lelli}, \citealt{Weldrake2003}, \cite{Swatersetal2003},  \cite{Swaters2011}. 

{\it (ii) Burkert profile:}\\ This (see eq. \ref{Burkert}) was utilized by \cite{Corbelli}, \cite{Shchelkanova}, \cite{CardoneDelPopolo2012}, and \cite {2009PhDT124F}. 

{\it (iii) Isothermal DM density distribution:} \begin{equation}    \rho_{iso}(r)=\frac{\rho_{0~iso}}{(1+(r/r_{0~iso})^2)^{3/2}}
\end{equation} (\citealt{Athanassoula}, \citealt{Chemin2006}, \citealt{Spano2008}).

{\it (iv) NFW profile:} \begin{equation}    \rho_{NFW}(r)=\frac{\rho_{s}}{(r/r_s)(1+(r/r_{s })^2)^{2}}
\end{equation} was adopted by \cite{Humphrey2006}, \cite{2009MNRAS393329N}, \cite{Napolitano2011}, \cite{2007astroph2233W}, \cite{2009MNRAS398561W},\cite{vandenBosch2001}.

To make our analysis more general, we transformed the DM
surface densities to the Burkert profile $\rho_0r_0$.
When the parameters
of the DM halo were given for pseudo-isothermal or NFW density
distributions, we translated them to the values of a Burkert DM halo
using the formulae from \cite{Boyarsky}: \begin{equation}    6.1r_{p.i.}=1.6r_0, 0.11 \rho_{0~p.i.}=0.37 \rho_0 \end{equation}    
\begin{equation}    r_{s}=1.6r_0, \rho_{s}=0.37 \rho_0. \end{equation}
For the case of an isothermal distribution, we used the transformation
from D09: \begin{equation} log(\mu_{0D})(Burkert)= log(\mu_{0D})(iso)+0.1.  \end{equation}

\begin{figure*}
\includegraphics[width=13cm,keepaspectratio]{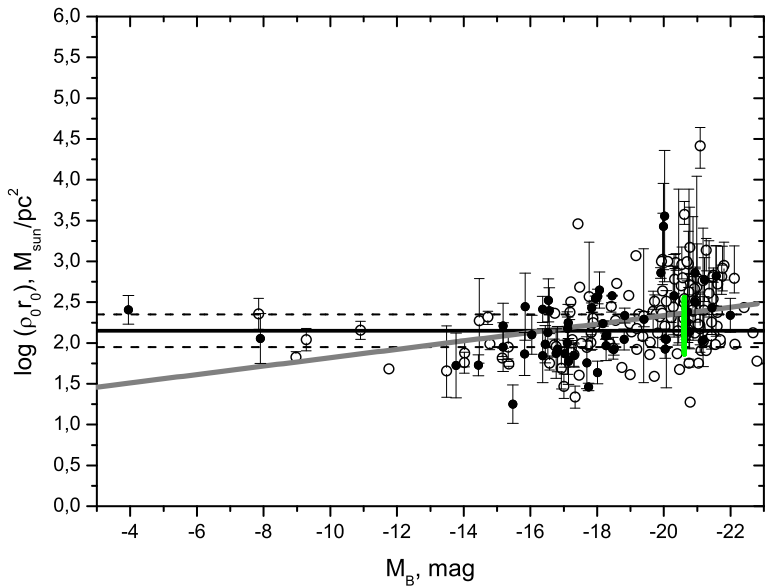}
\caption{``$\log(\rho_0r_0)$ -- $M_B$'' diagram. 
Solid and dashed black lines represent the value of $\log \mu_{0D} =2.15\pm 0.2$ obtained by 
Donato et al. (2009).
%D09. 
Gray line denotes the linear regression. Circles correspond to the DM surface densities obtained for the sample. Galaxies for which the DM parameters were obtained more than in one source and the averaged estimates of $\log(\rho_0r_0)$  were made are marked by filled symbols. Green line denotes the range of different dark halo parameters for UGC5175.}
\label{mu0_mb}
\end{figure*}

We compared the dark halo surface density with the absolute B
magnitude, as was done by D09 (see Fig. \ref{mu0_mb}). Circles correspond 
to the DM surface densities obtained for the sample. Galaxies for
which DM parameters were obtained in more than one source and
averaged estimates of the DM parameters obtained are marked by
filled symbols. The error bars in Fig. \ref{mu0_mb} are calculated from the
error values of the DM halo parameters given in the source article,
according to the formula \begin{equation} \Delta f(p_1,p_2)=((\partial f/ \partial p_2\Delta p_2)^2+ (\partial f/ \partial p_1\Delta p_1)^2)^{0.5}. \end{equation}

% In \cite{THINGS} and \cite{Lelli} the errors of DM parameters are related to the uncertainties of the %rotation curves; \cite{Barnes} calculated the uncertainties in each fitted parameter using a Markov %Chain Monte Carlo method; in \cite{Saburova2012} the error bars correspond to the uncertainty of %the rotation curve and stellar velocity dispersion. 

\begin{figure*}
\includegraphics[width=11cm,keepaspectratio]{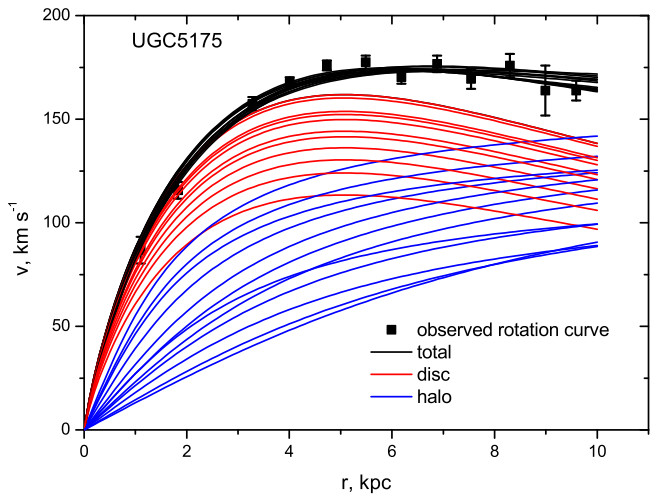}
\caption{The different variants of decomposition of the rotation curve of UGC5175 taken from Spano et al. (2008) (squares with error bars). Blue and red lines denote the DM halo and disc contributions to the rotation curve obtained in different models. Black lines correspond to different model rotation curves.}
\label{rcdecomposition}
\end{figure*}

To define the typical error bar that occurs due to the uncertainty in
the rotation curve decomposition, we constructed different models
of the rotation curve of UGC 5175 (as an example), using a fixed
scalelength of baryonic matter $h_d=2.36$ kpc (\citealt{Spano2008}). We
used a pseudo-isothermal density distribution of the DM halo. The
rotation curve was taken from \cite{Spano2008}. The parameters
obtained for the DM halo and stellar disc of the galaxy are given in Table \ref{parameters}, while the corresponding rotation curve models are shown
in Fig. \ref{rcdecomposition}. Blue, red, and black lines in Fig. \ref{rcdecomposition} correspond to the 
contributions of the DM halo and the disc to the rotation curve and
the total model rotation curve.
\begin{table*}
\begin{center}
\caption{ DM halo and disc parameters for different models of the rotation curve of UGC5175. (1) -- the number of model; (2) -- the core radius; (3) -- the DM halo central density; (4) -- the DM halo surface density translated to the Burkert density distribution; (5) -- exponential disc central surface density; (6) -- the reduced  $\chi^2_r$ \label{parameters}}

    \begin{tabular}{llllll}
    \hline
Number&$r_0$&$\rho_0$&$log(\mu_{0D})$&$\sigma_{0d}$&$\chi^2_r$\\
&kpc&$10^{-3}\cdot M_{\odot}/pc^3$&$M_{\odot}/pc^2$&$M_{\odot}/pc^2$&\\
\hline
 \hline
(1)&(2)&(3)&(4)&(5)&(6)\\
 \hline
1	&	2.00	&	98.28	&	2.35	&	747	&	1.65	\\
2	&	3.54	&	31.32	&	2.10	&	904	&	1.89	\\
3	&	6.91	&	9.58	&	1.88	&	1055	&	1.87	\\
4	&	5.37	&	11.94	&	1.86	&	1055	&	1.70	\\
5	&	3.78	&	18.89	&	1.91	&	1034	&	1.59	\\
6	&	3.06	&	43.41	&	2.18	&	837	&	2.36	\\
7	&	1.49	&	213.31	&	2.56	&	518	&	3.13	\\
8	&	1.59	&	164.60	&	2.47	&	620	&	2.56	\\
9	&	1.60	&	146.84	&	2.43	&	684	&	2.11	\\
10	&	2.46	&	44.90	&	2.10	&	934	&	1.47	\\
11	&	3.45	&	26.77	&	2.02	&	952	&	1.73	\\
12	&	2.60	&	60.49	&	2.25	&	807	&	1.80	\\
\hline

 \end{tabular}
\end{center}
\end{table*}

  The range of different dark halo parameters for UGC 5175 is
shown by the green line in Fig. \ref{mu0_mb} and below; this line could be interpreted as
the typical error bar for objects with absent error values.

From Fig. \ref{mu0_mb}, it can be seen that the scatter of the dark matter halo
parameters is twice as high as that found by D09. The mean value
is $\log \mu_{0D} =2.26\pm0.46 M_{\odot} pc^{-2}$. A very slight trend can also
be seen towards higher values of $\log(\mu_{0D}) $ for more luminous (and
consequently more massive) systems, in agreement with what was
previously found by \cite{Boyarsky}. \cite{Boyarsky} claimed that this trend is in good agreement with cosmological simulations
of dark matter by \cite{2008MNRAS3911940M}. The linear regression line is shown in Fig. \ref{mu0_mb} by gray line\footnote{Here and below we calculated a simple linear regression.}. The linear regression equation can be found in Table \ref{tab1}. The correlation coefficient $R$ of the trend is quite low; however, it is still statistically significant. Table \ref{tab1} gives also the twice standard error of correlation coefficient $2\sigma_R=2((1-R^2)/(N-2))^{0.5}$, where N is the number of objects. The correlation is also significant if we consider only
those galaxies for which several DM parameter estimations were available and averaged (see Table \ref{tab1}). The averaged estimates are
more reliable, since the systematic effects of the various approaches
to DM parameter evaluation have less influence.

\begin{figure*}
\includegraphics[width=13cm,keepaspectratio]{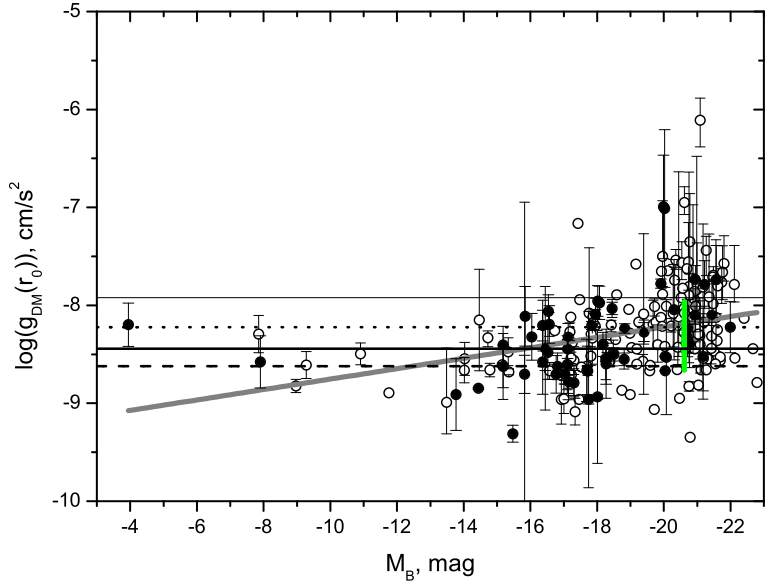}
\caption{``$\log(g_{DM}) $ -- $M_B$'' diagram. Gray line denotes the linear regression.  Circles correspond to the DM Newtonian accelerations obtained for the sample. Galaxies for which the DM parameters were obtained more than in one source and the averaged estimates of $g_{DM}$  were made are marked by filled symbols. Green line shows the range of different dark halo parameters for UGC5175. Black horizontal lines correspond to the halo maximum accelerations predicted by MOND (see the text): $\log(0.2a_0)$ (dashed line), $\log(0.3a_0)$ (thick line), $\log(0.4a_0)$ (dotted line), and $\log(a_0)$ (thin line).}
\label{gdm_mb}
\end{figure*}

Besides the $\log(\rho_0r_0)$ -- $M_B$ diagram, we also compared the DM Newtonian acceleration within $r_0$, namely $g_{DM}=GM_{DM}/r_0^2$, with the  absolute B-band magnitude (see Fig. \ref{gdm_mb}).

  From Fig. \ref{gdm_mb}, it follows that there is a slight trend towards higher
accelerations for higher luminosities, in good agreement with the
findings of \cite{CardoneDelPopolo2012}. Our sample includes the sample of \cite{CardoneDelPopolo2012} and is larger than it. Additionally, we utilize different techniques of DM parameter
evaluation, while \cite{CardoneDelPopolo2012} used only best-fit modelling. The gray line in Fig. \ref{gdm_mb} denotes the linear regression
line. The equation and correlation coefficient with twice its standard
error are given in Table \ref{tab1} for both the entire sample and only those
galaxies with averaged estimates. 

From Figs. \ref{mu0_mb} and \ref{gdm_mb}, it can be seen that the behaviour of high and
low-luminosity systems in the diagrams is different. Dwarf
galaxies with $M_B>-12$ mag seem to have lower variation of
DM surface density and acceleration with absolute magnitude, in
contrast to high-luminosity systems, which show a stronger linear
trend of these DM parameters with luminosity. To study this
dichotomy, we divided the data into bins by the absolute B-band
magnitude. Each bin had a width of 0.5 mag. After that, we calculated
the median values of DM surface densities and accelerations for each bin and found a linear regression equation for the relation
between the median values of DM parameters and absolute magnitudes.
This approach allowed us to give more weight to the less
numerous dwarf galaxies. The correlations become much weaker
for the median fit: $\log(\rho_0 r_0)=(1.83 \pm 0.16)-(0.02 \pm 0.01) \cdot M_B$; ($R\pm 2\sigma=0.34\pm 0.38$) and $\log(g_{DM}(r_0))=(-8.82\pm0.16)-(0.02\pm 0.01)\cdot M_B$; ($R\pm 2\sigma=0.43\pm 0.37$). However, if we don't take the dwarf galaxies ($M_B>-12$ mag) into account, the situation changes and the correlations become more significant: $\log(\rho_0 r_0)=(1.00 \pm 0.24)-(0.06 \pm 0.01) \cdot M_B$; ($R\pm 2\sigma=0.75\pm 0.31$) and $\log(g_{DM}(r_0))=(-9.56\pm0.26)-(0.06\pm 0.01)\cdot M_B$; ($R\pm 2\sigma=0.72\pm 0.33$). This analysis confirms
the dichotomy of low- and high-luminosity galaxies in the diagrams;
however, to make further conclusions one needs more data on dwarf
galaxies. A similar dichotomy was previously found by \cite{Napolitano2010}, who compared the projected density of DM in the
central regions of galaxies with stellar masses for a large sample
of galaxies. According to this comparison, galaxies with low stellar
masses have almost universal projected DM density, while for other
objects there is a trend of DM surface density with stellar mass.  

Since the universality of DM surface density is in good agreement
with MOND (see \citealt{Milgrom2009}), we decided to study how
our results correspond to MOND. \cite{Milgrom2009} concluded that
MOND predicts a quasi-universal value of $log(\mu_{0D})=2.14$, which
is very close to the value obtained by D09, shown in Fig. \ref{mu0_mb} by a thick black line. According to \cite{Milgrom2009}, the quasi-universal
value is not shared by objects with low surface densities, so the DM
surface density could be lower. Thus, we should expect the DM surface
density to be around 2.14 or lower. As can be seen from Fig. \ref{mu0_mb} this prediction is not confirmed by our data for high-luminosity
systems.

\cite{MilgromSanders2005} have shown that MOND predicts the
existence of the maximum possible halo acceleration, which spans
a range of $0.2a_0-0.4a_0$, where $a_0$ is the MOND constant. We have overplotted in Fig. \ref{gdm_mb} the values of $\log(0.2a_0)$ (dashed black line), $\log(0.3a_0)$ (thick black line), $\log(0.4a_0)$ (dotted black line), and $\log(a_0)$ (thin black line). From Fig. \ref{gdm_mb} it can be seen that our data do
not confirm this prediction of MOND. Some of the galaxies have
values of $g_{DM}(r_0)$ which are higher than not only $0.4a_0$ but even $a_0$.

\begin{figure*}
\includegraphics[width=13cm,keepaspectratio]{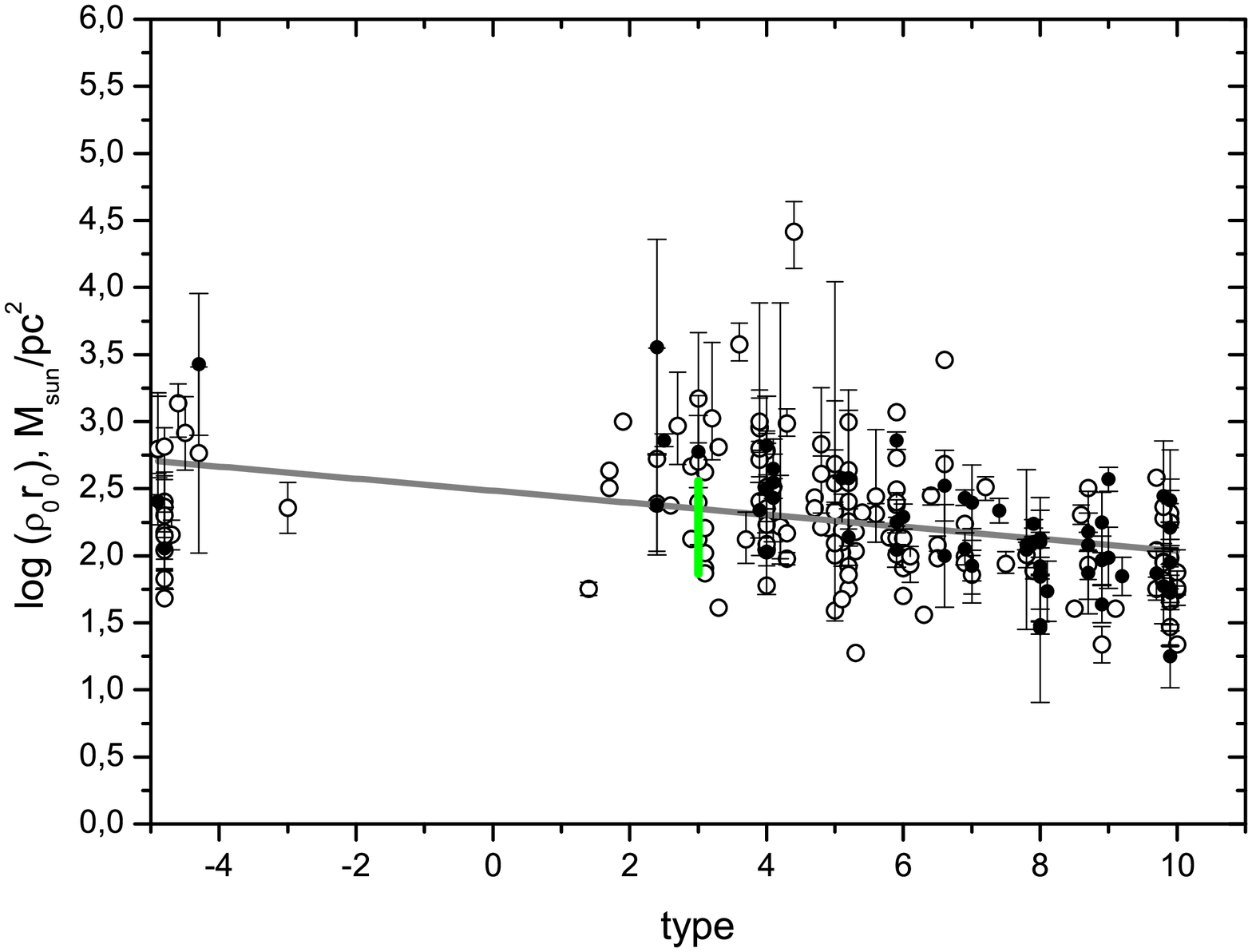}
\caption{``$\log(\rho_0r_0)$ -- morphological type '' diagram. Gray line denotes the linear regression.  Circles correspond to the DM surface densities obtained for the sample. Galaxies for which the DM parameters were obtained more than in one source and the averaged estimates of $\log(\rho_0r_0)$  were made are marked by filled symbols. Green line shows the range of different dark halo parameters for UGC5175.}
\label{mu0_t}
\end{figure*}

\begin{figure*}
\includegraphics[width=13cm,keepaspectratio]{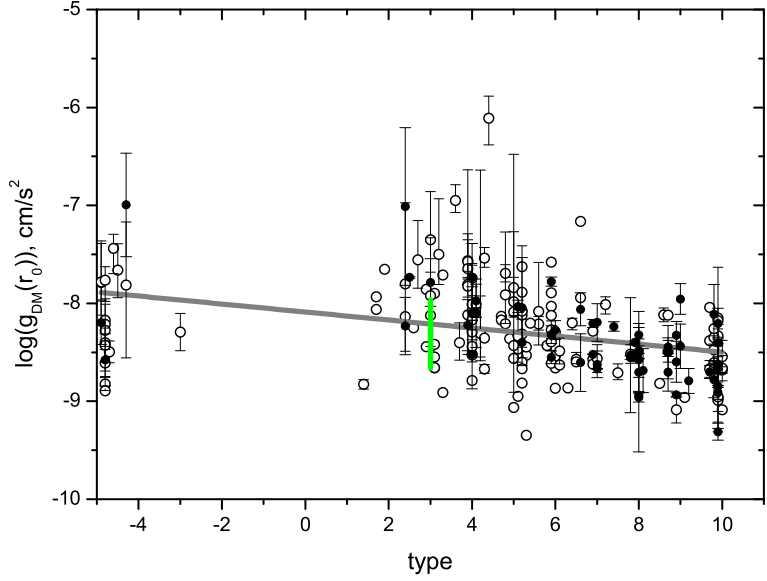}
\caption{``$\log(g_{DM}) $ -- morphological type'' diagram. Circles correspond to the DM accelerations obtained for the sample. Galaxies for which the DM parameters were obtained more than in one source and the averaged estimates of acceleration  were made are marked by filled symbols. Green line shows the range of different dark halo parameters for UGC5175.  }
\label{gdm_t}
\end{figure*}

\begin{figure*}
\includegraphics[width=13cm,keepaspectratio]{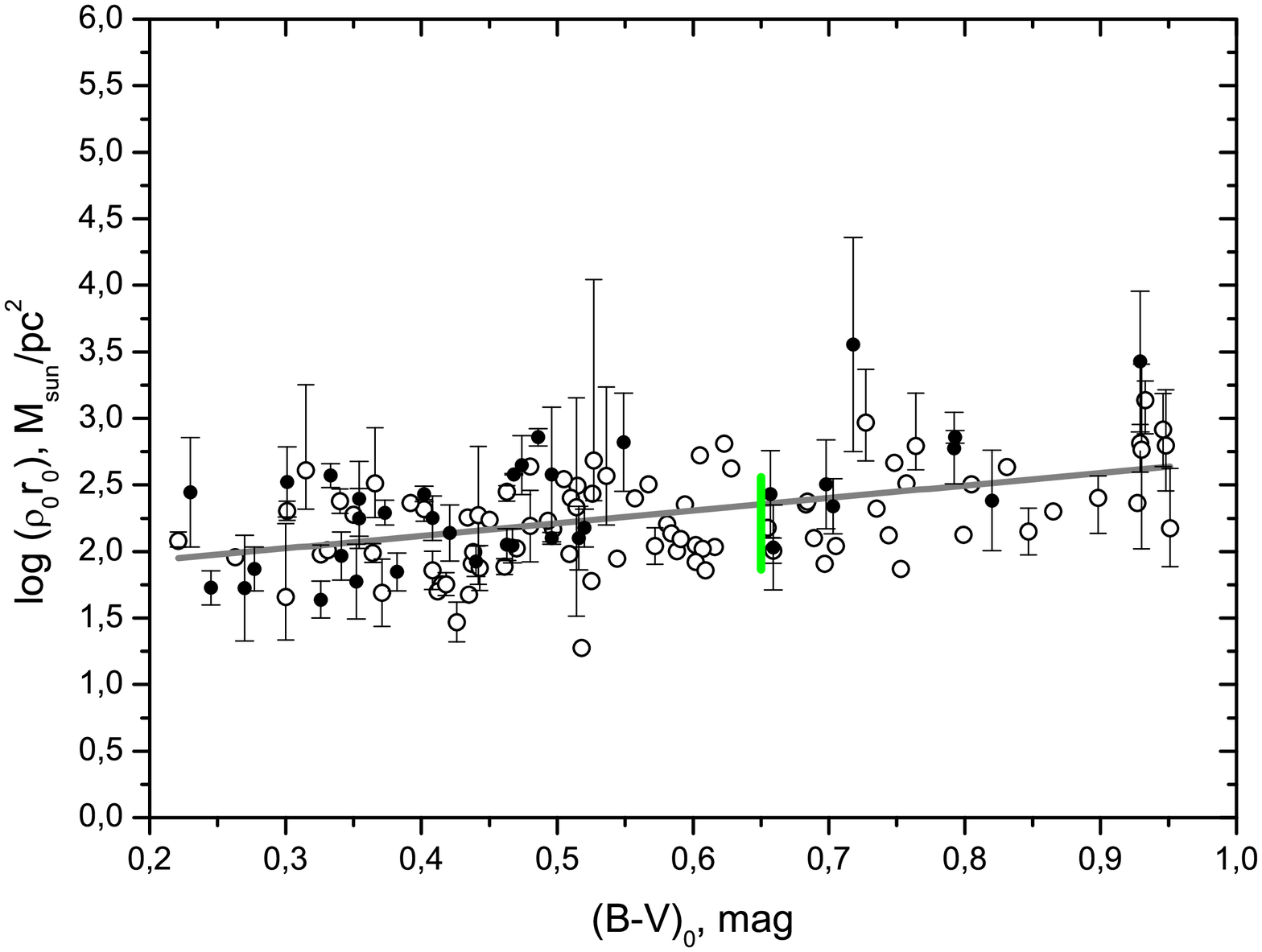}
\caption{``$\log(\rho_0r_0)$ -- $(B-V)_0$ '' diagram. 
Gray line denotes the linear regression. Circles correspond to the DM surface densities obtained for the sample. Galaxies for which the DM parameters were obtained more than in one source and the averaged estimates of $\log(\rho_0r_0)$  were made are marked by filled symbols. Green line shows the range of different dark halo parameters for UGC5175.}
\label{mu0_bv}
\end{figure*}

\begin{figure*}
\includegraphics[width=13cm,keepaspectratio]{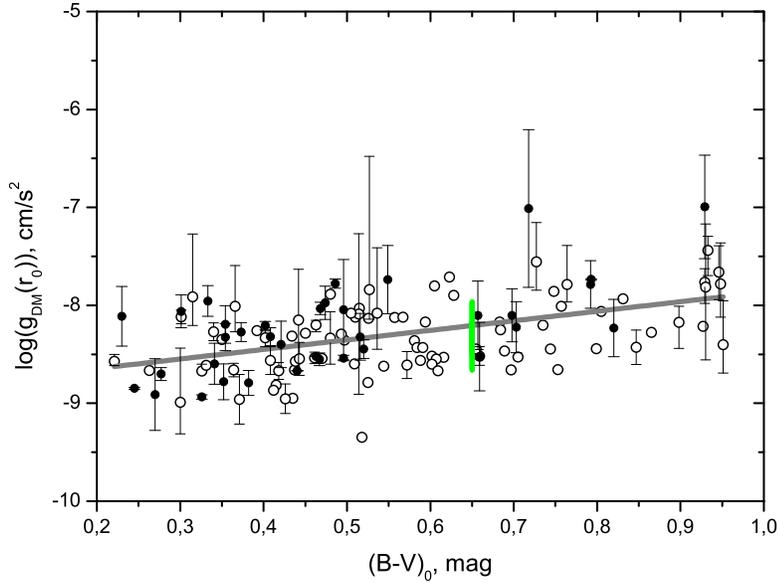}
\caption{``$\log(g_{DM}) $ -- $(B-V)_0$'' diagram. Circles correspond to the DM accelerations obtained for the sample. Galaxies for which the DM parameters were obtained more than in one source and the averaged estimates of $g_{DM}$  were made are marked by filled symbols. Green line shows the range of different dark halo parameters for UGC5175. Gray line denotes the linear regression. }
\label{gdm_bv}
\end{figure*}

\begin{figure*}
\includegraphics[width=13cm,keepaspectratio]{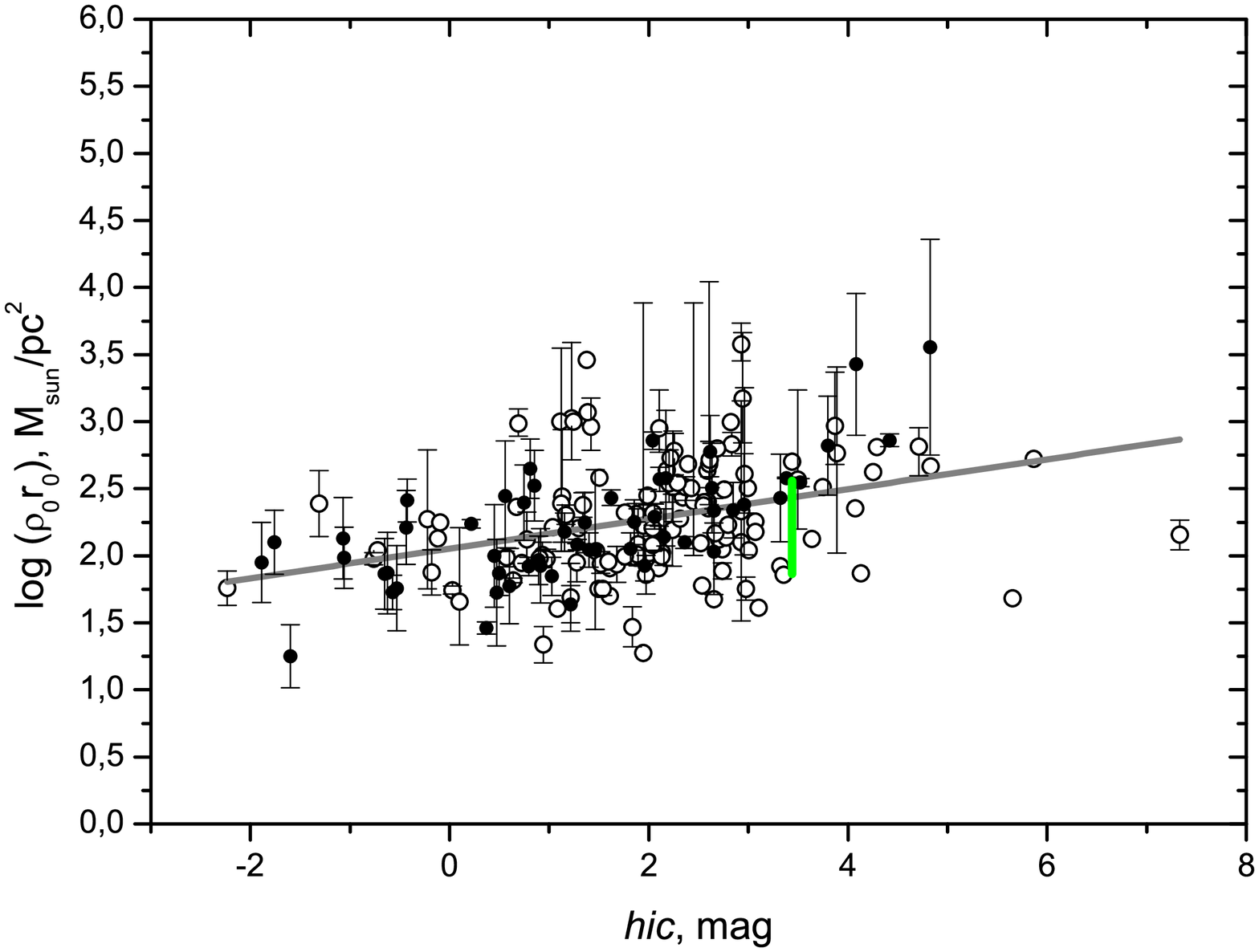}
\caption{``$\log(\rho_0r_0)$ -- hydrogen content per B-band flux $hic$ '' diagram.  Gray line denotes the linear regression. Circles correspond to the DM surface densities obtained for the sample. Galaxies for which the DM parameters were obtained more than in one source and the averaged estimates of $\log(\rho_0r_0)$  were made are marked by filled symbols. Green line shows the range of different dark halo parameters for UGC5175. }
\label{mu0_hic}
\end{figure*}

\begin{figure*}
\includegraphics[width=13cm,keepaspectratio]{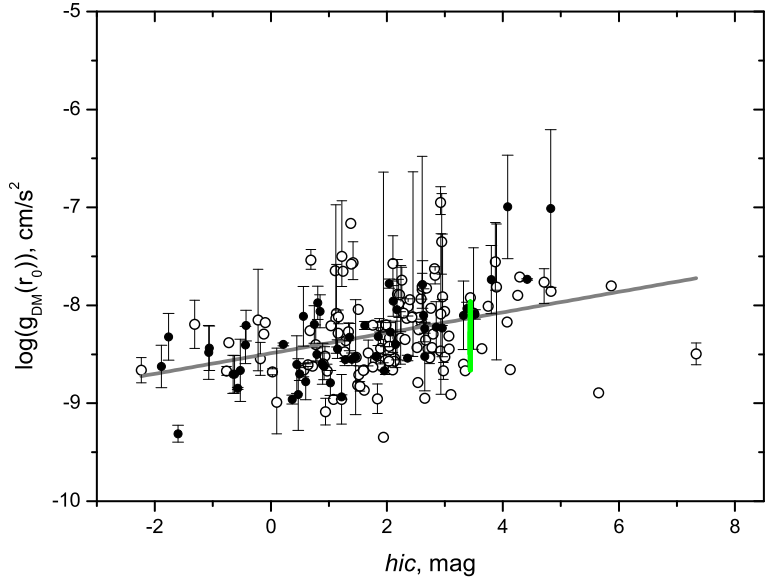}
\caption{``$\log(g_{DM}) $ -- hydrogen content per B-band flux $hic$ '' diagram. Gray line denotes the linear regression. Circles correspond to the DM accelerations obtained for the sample. Galaxies for which the DM parameters were obtained more than in one source and the averaged estimates of $g_{DM}$  were made are marked by filled symbols. Green line shows the range of different dark halo parameters for UGC5175.}
\label{gdm_hic}
\end{figure*}

The relation between the parameters of the DM halo and the total
B-band luminosities of the galaxies indicates that DM haloes could
play a significant role in the processes of formation and evolution of
galaxies. In order to make further progress, we decided to compare
the surface density of the DM halo and its Newtonian acceleration
with the observational properties of galaxies, such as morphological
type, colour index and content of neutral hydrogen. All these
quantities were found in the Hyperleda catalogue \footnote{http://leda.univ-lyon1.fr/} (\citealt{ Paturel2003}) for most galaxies of the sample. In Figs. \ref{mu0_t}, \ref{gdm_t} we compare the surface density and 
Newtonian acceleration of DM haloes with the morphological types
of galaxies. From these figures, it is evident that the parameters of
DM haloes correlate with galaxy type. The correlation coefficients
with their standard errors and the linear regression equation can
be found in Table \ref{tab1}, both for the entire sample and for only those
galaxies with mean DM parameters. From Figs \ref{mu0_t} and \ref{gdm_t}, it follows
that galaxies with earlier morphological types, possessing a higher contribution of the bulge to the luminosity, tend to have higher values
of DM halo surface densities $\mu_{0D} $ and accelerations $g_{DM}(r_0)$ in
comparison with late-type galaxies. Weak but statistically significant
correlations are also found between DM halo surface density
and acceleration and the colour indices $(B-V)_0$ corrected for dust
extinction (see Figs. \ref{mu0_bv} and \ref{gdm_bv}). We also found a very slight but significant
correlation between the DM halo parameters and the hydrogen
content per unit of B-band flux, $hic=m_{21cm}-m_B$ (Figs. \ref{mu0_hic}, \ref{gdm_hic}). Galaxies with higher contents of hydrogen possess lower values
of $hic$.   The correlation coefficients and the linear regression coefficients
for the relations are given in Table \ref{tab1}. As can be seen from
Table \ref{tab1}, the correlation coefficients are higher for galaxies with
averaged DM parameter estimates. 

The correlations found could mean that early-type galaxies with
passive star formation and low content of gas have denser DM
haloes than late-type ones. If these correlations are real, then one
can conclude that the properties of the DM halo could determine
the evolution of the visible galaxy. Galaxies with denser DM haloes
could form their stars more effectively and lose their gas faster
than those with less massive haloes. Also, one cannot exclude the possibility that the interaction between DM and baryons had an
influence on the DM density distribution (see, e.g. \citealt{Wu2014}).

\begin{figure*}
\includegraphics[width=13cm,keepaspectratio]{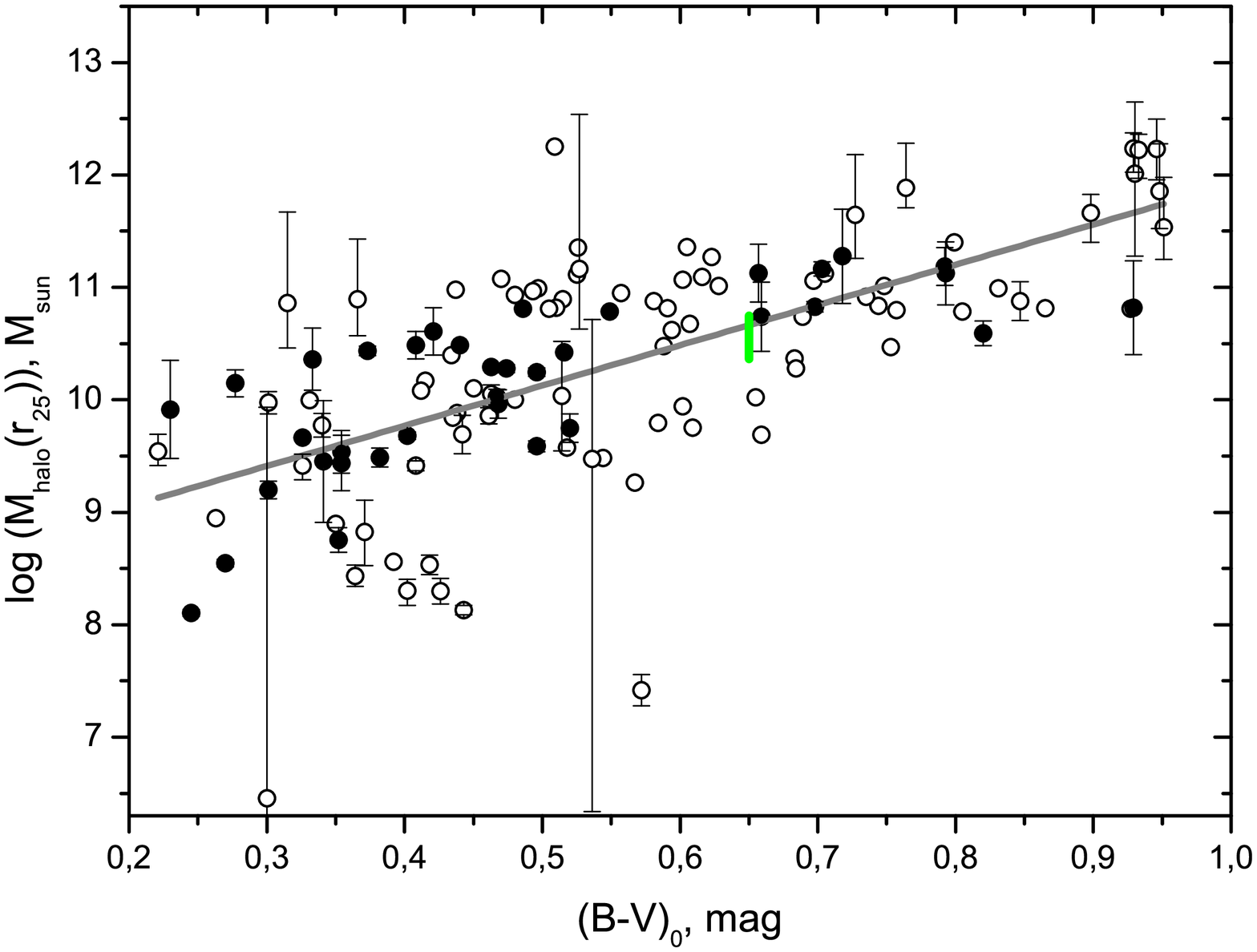}
\caption{The logarithm of the DM halo mass within the optical radius $\log(M_{halo}(r_{25}))$ compared to the $(B-V)_0$ colour index.
Gray line denotes the linear regression. Galaxies for which the DM parameters were obtained more than in one source and the averaged estimates of the DM halo masses were made are marked by filled symbols. Green line shows the range of different dark halo masses for
UGC5175. }
\label{mh_bv}
\end{figure*}

The correlations between DM halo central surface density $\mu_{0D}$ and acceleration $g_{DM}$ and $(B-V)_0$ could be the result of the relation
between DM halo mass and the colour of a galaxy. According
to cosmological simulations, red galaxies reside in more
massive DM haloes in comparison with blue ones (see e.g. \citealt{Wang2012}, \citealt{Guo2011}). To test whether our results agree
with this conclusion, we compared the DM halo masses inside optical
radius $r_{25}$ and the $(B-V)_0$ colour indices (see Fig. \ref{mh_bv}). For
high-surface-brightness galaxies, we used the radius of the B-band 25 $ mag ~arcsec^{-2}$ isophote as the optical radius, while for LSBs we
utilized four exponential disc scalelengths instead (the scalelengths
were taken from the source articles). From Fig. \ref{mh_bv}, it follows that the
correlation between DM halo mass and $(B-V)_0$ colour is stronger than that for DM surface density and acceleration (see Table \ref{tab1} for the correlation coefficients). This could mean that the relation between
the surface density and Newtonian acceleration of the DM
halo and the colour index is the manifestation of a stronger link
between DM halo mass and the evolutionary state of a galaxy.

 Another thing that could follow from the correlations is that
red galaxies could undergo more merging events in comparison
with blue ones. We tested whether isolated galaxies differ from
other galaxies of the sample in their values of $\mu_{0D}$. To do this,
we selected the galaxies of our sample presented in the Analysis of
the interstellar Medium of Isolated GAlaxies (AMIGA) catalogue of
isolated systems\footnote{http://amiga.iaa.es/p/1-amiga-home.htm}. The catalogue provides a list of galaxies that have
not interacted with a significant mass neighbour in the past $\sim$ 3 Gyr. The mean value of the DM surface density of isolated galaxies $\log(\mu_{0D})=2.35\pm0.42$ does not differ from that for the entire
sample (see above). According to the Kolmogorov-Smirnov test at
0.05 level, these galaxies are not significantly different from the
whole sample.



\begin{table*}
\begin{center}
\caption{The linear regression equations. (1) -- the linear regression equation; (2) -- the correlation coefficient $R$; (3) -- standard error of the correlation coefficient multiplied by 2 $2\sigma_R$. \label{tab1}}

 \begin{tabular}{lll}
    \hline
Equation&$R$&$2\sigma_R$\\
\hline
 \hline
(1)&(2)&(3)\\
\hline
 \hline
The entire sample&&\\
$\log(\rho_0 r_0)=(1.31 \pm 0.2)-(0.05 \pm 0.01) \cdot M_B$&0.32&0.13\\
$\log(g_{DM}(r_0))=(-9.28\pm0.20)-(0.05\pm 0.01)\cdot M_B$&0.33&0.13\\
\hline
Galaxies with the averaged estimates&&\\
$\log(\rho_0 r_0)=(1.45 \pm 0.32)-(0.04 \pm 0.02) \cdot M_B$&0.28&0.26\\
$\log(g_{DM}(r_0))=(-9.15\pm0.31)-(0.04\pm 0.02)\cdot M_B$&0.30&0.26\\
\hline
\hline
The entire sample&&\\

$\log(\rho_0 r_0)=(2.48 \pm 0.05)-(0.04 \pm 0.01) \cdot t$&0.37&0.13\\
$\log(g_{DM}(r_0))=(-8.09\pm0.05)-(0.04\pm 0.01)\cdot t$&0.34&0.13\\
\hline
Galaxies with the averaged estimates&&\\
$\log(\rho_0 r_0)=(2.67 \pm 0.10)-(0.07 \pm 0.01) \cdot t$&0.56&0.22\\
$\log(g_{DM}(r_0))=(-7.90\pm0.10)-(0.07\pm 0.01)\cdot t$&0.56&0.22\\
\hline
\hline
The entire sample&&\\

$\log(\rho_0 r_0)=(1.73 \pm 0.10)+(0.96 \pm 0.17) \cdot (B-V)_0$&0.46&0.16\\
$\log(g_{DM}(r_0))=(-8.85\pm0.10)+(0.99\pm 0.17)\cdot (B-V)_0$&0.46&0.16\\
\hline
Galaxies with the averaged estimates&&\\
$\log(\rho_0 r_0)=(1.59 \pm 0.17)+(1.52 \pm 0.33) \cdot (B-V)_0$&0.60&0.27\\
$\log(g_{DM}(r_0))=(-9.01\pm0.17)+(1.59\pm 0.34)\cdot (B-V)_0$&0.61&0.27\\
\hline
\hline
The entire sample&&\\

$\log(\rho_0 r_0)=(2.05 \pm 0.05)+(0.11 \pm 0.02) \cdot hic$&0.38&0.14\\
$\log(g_{DM}(r_0))=(-8.49\pm0.05)+(0.10\pm 0.02)\cdot hic$&0.36&0.14\\
\hline
Galaxies with the averaged estimates&&\\
$\log(\rho_0 r_0)=(1.98 \pm 0.05)+(0.19 \pm 0.03) \cdot hic$&0.68&0.22\\
$\log(g_{DM}(r_0))=(-8.58\pm0.06)+(0.18\pm 0.03)\cdot hic$&0.67&0.21\\
\hline
\hline
The entire sample&&\\

$\log(M_{halo}(r_{25}))=(8.34\pm0.22)+(3.58\pm 0.38)\cdot (B-V)_0$&0.65&0.14\\
\hline
Galaxies with the averaged estimates&&\\
$\log(M_{halo}(r_{25}))=(8.60\pm0.26)+(3.18\pm 0.49)\cdot (B-V)_0$&0.73&0.23\\
\hline
\hline 
 \end{tabular}
 
 \end{center}
\end{table*}

\section{Discussion and Conclusions}

The galaxy formation process is a complex topic not completely
understood to date. Scaling relations can help us to extend our
knowledge on this topic. When certain parameters of galaxies turn
out to be universal, they have great importance, since universality
implies the existence of some important physical mechanism
driving the process studied. In our case, the universality of DM
surface density described in D09 and G09 has a strong meaning. It
strongly couples DM and baryon evolution in galaxies (see G09):
a large central luminous density hints at a large core, which means
a small quantity of DM in the central part of the galaxy. This low
quantity of DM in the centre of dwarfs has been related by G09
to the ``mass discrepancy-acceleration relation" (\citealt{McGaugh2004}; \citealt{1983ApJ270365M}), and MOND. In more standard scenarios, which
are based on the idea that the $\Lambda$CDM model is able to explain the
universe on small scales when baryon physics is taken appropriately
into account, the low DM density in dwarf cores has been
explained as a by-product of supernova explosions (\citealt{1996MNRAS283L72N}; \citealt{Gelato1999}; \citealt{Read2005}; \citealt{Mashchenko2006}; \citealt{Governato2010}) or resulting from
the transfer of energy and angular momentum from baryons to DM
through dynamical friction (\citealt{El-Zant2004, El-Zant2001}; \citealt{2008ApJ685L105R}; \citealt{DelPopolo2009}; \citealt{Cole2011}).
%, more standard explanations, which are based on the idea that the $\Lambda$CDM model is able to explain the universe on small scales when baryon physics is taken appropriately into account. 

Motivated by the results of D09 and G09, in the present article
we revisited the constancy (universality) of the  DM halo surface
density $\mu_{0D}$ and its Newtonian acceleration $g_{DM}$ within the core
radius for galaxies of different luminosities found by D09. The
adopted sample is different from that of D09 and we used DM
parameters obtained using some methods different from that used
in D09. According to our study, the dispersion of values of surface
density $\mu_{0D}$ is twice as high as that of D09 and G09. Also, we
found that both $\mu_{0D}$ and $g_{DM}$ correlate with luminosity, in good
agreement with the previous findings of \cite{CardoneDelPopolo2012} and \cite{Boyarsky}. We also found that galaxies with
high and low luminosities behave differently in $\mu_{0D}$ and $g_{DM}$ versus $M_B$ diagrams. Faint objects (with $M_B>-12$ mag) have low variation
of DM surface densities and accelerations with absolute B-band
magnitude, in contrast to luminous systems, which tend to have a
strong relation between these DM properties and luminosity.

We also revealed weak but statistically meaningful correlations
between these parameters and the morphological type, colour index
and content of neutral hydrogen per B-band flux. Galaxies of early
types and with low content of gas tend to have higher values of $\mu_{0D}$ and  $g_{DM}$. A relation between DM halo surface density and
acceleration and $(B-V)_0$ colour index seems to be a manifestation
of a stronger correlation with DM halo mass. This conclusion is
in agreement with the results of cosmological simulations, where
red galaxies reside in more massive DM haloes in comparison with
blue ones (see e.g. \citealt{Wang2012}, \citealt{Guo2011}). 
The
correlations are statistically significant for both the entire sample
of galaxies and the most reliable averaged DM parameter estimates
for galaxies with several evaluations of the DM parameters. This
could indicate that intrinsic systematic effects of the methods of
evaluation of the DM density distribution may have insignificant
influence on our results.

We also found that the difference between galaxies with high
and low values of $\mu_{0D}$ is not due to merging events in the past 3 Gyr, since the DM surface densities of isolated AMIGA galaxies
of our sample are not significantly different from those of the entire
sample.

Our results contradict the universality of DM density and acceleration
found by D09 and G09. The quoted constancy of surface
density and Newtonian acceleration in $r_0$ is one of MOND's prediction. Rephrasing Milgrom (2009), the surface density is constant
for objects of any internal structure and mass, if they are in the
Newtonian regime (see Milgrom 2009). This MOND prediction is
not confirmed by our data. Our result is also at odds with the MOND
prediction of maximal halo acceleration (see \citealt{MilgromSanders2005}).

Thus, our results seem to contradict the conclusions of D09 and
G09. At the same time, they do not confirm the MOND predictions
related to the inner structure of galaxies.

%The previous discussion puts obvious doubts on the D09 and G09 conclusions, based on the assumption that all the galaxies in %their sample are well fitted by cored profiles, namely the Burkert profile. 

%
%%the evolution of visible galaxies depends on the properties of DM halo. Galaxies with denser DM haloes transform their gas into %%stars more effectively than those with less massive haloes. 
%
{\bf Acknowledgments} We thank the anonymous referee for the important remarks which allowed us to improve the paper. A.S. thanks Kirill Zaslavskiy for helpful comments and the support. A.D.P. would like to thank the International Institute of Physics in Natal for the facilities and
hospitality. We thank Charles Downing from Exeter University for a critical reading
of the paper. We are grateful to Anatoly Zasov for fruitful discussion. We wish to thank Anastasia Kasparova for providing us the DM parameters for the sample of galaxies from Kasparova 2012. This publication makes use of the AMIGA VO-archive, supported by Grant AYA2008-06181-C02 and AYA2011-30491-C02-01, co-financed by MICINN and FEDER funds, and the Junta de Andalucía (Spain) grant P08-FQM-4205. We acknowledge the usage of the HyperLeda database (http://leda.univ-lyon1.fr). This work was partly supported by the Russian Foundation for Basic Research (RFBR), grants no.  14-22-03006 and 12-02-00685-a. 

\bibliographystyle{mn2e}
\bibliography{saburova}

\onecolumn
\appendix
\section{The sample}
\begin{longtable}{ccccccccc}
\caption{The properties of sample galaxies. (1) -- the name of the galaxy; (2) -- the adopted distance; (3) -- morphological type taken from Hyperleda database; (4) -- absolute B-band magnitude; (5) -- logarithm of DM halo central surface density; (6) -- logarithm of the DM Newtonian acceleration at $r_0$; (7) -- logarithm of the DM halo mass inside the optical radius; (8) -- the source of the DM halo parameters:  1-- Athanassoula et al. (1987), 2 --  Barnes et al. (2004), 3 -- Begum \& Chengalur (2004), 4 – Chemin et al. (2006), 5 -- Corbelli (2003), 6 -- de Blok \& Bosma (2002),  7 -- Cardone \& Del Popolo (2012), 8 -- Lelli et al. (2010), 9 – Humphrey et al. (2006), 10 -- Kasparova (2012), 11 -- Kuzio de Naray et al. (2008), 12 -- Frigerio Martins (2009), 13 -- de Blok \& McGaugh (1997), 14 -- Napolitano et al. (2009), 15 -- Napolitano et al. (2011), 16 -- Saburova \& Zasov (2012), 17 -- Shchelkanova et al. (2013), 18 -- Spano et al. (2008), 19 – Swaters et al. (2011), 20 -- Swaters et al. (2003), 21 --  de Blok et al. (2008), 22 -- van den Bosch \& Swaters (2001), 23 -- van Eymeren et al.  (2009), 24 -- Weijmans  et al. (2008), 25 -- Weijmans  et al. (2009), 26 – Weldrake et al. (2003), 27 – Wu (2007), 28 -- Kasparova et al. (2014); (9) – the method of estimation of the DM parameters: I -- best fit model, II -- maximum disc model, III –the method based on the stellar population synthesis model colour-mass-to-light ratio relation, IV -- the approach based on the spectral energy distribution and long-slit spectra modeling, V -- the method relying on the marginal gravitational stability of the disc, VI -- the approach taking into account the presence of spiral structure, VII -- the method based on X-ray properties of
hot gas VIII -- Jeans analysis based on planetary nebulae or stellar kinematical data, IX -- the modeling reproducing the observed velocity dispersion profiles and number density profiles, X --  "Bottema disc" of  de Blok \& McGaugh (1997).  } \label{sample}\\
%\begin{supertabular}{lllllllll}
 %\begin{tabular}{lllllllll}
   \hline Galaxy&$D$&Type & $M_B$ &$\log(\mu_{0D})$&$\log(g_{DM})$&$\log(M_{halo}(r_{25}))$&ref.&method\\
&Mpc&&mag.&$M_{\odot}~pc^{-2}$&$cm~s^{-2}$ & $M_{\odot}$&&\\\hline
 \hline
(1)&(2)&(3)&(4)&(5)&(6)&(7)&(8)&(9)\\
 \hline
\endfirsthead

\caption{ Continued }\label{sample} \\
\hline
\hline (1)&(2)&(3)&(4)&(5)&(6)&(7)&(8)\\
\hline 
   \hline Galaxy&$D$&Type & $M_B$ &$\log(\mu_{0D})$&$\log(g_{DM})$&$\log(M_{halo}(r_{25}))$&ref.&method\\
&Mpc&&mag.&$M_{\odot}~pc^{-2}$&$cm~s^{-2}$ & $M_{\odot}$&&\\
\hline
 \hline
(1)&(2)&(3)&(4)&(5)&(6)&(7)&(8)&(9)\\
 \hline
\endhead
  ESO059-001     	&	4.57	&	 IB              	&	-15.33	&	1.95	$^	{+	0.14	}	_	{-	0.14	}$	&	-8.47	$^	{+	0.14	}	_	{-	0.14	}$	&	8.34	$^	{+	0.09	}	_	{-	0.14	}$	&	23	&	I	\\
  ESO186-055     	&	60.10	&	 Sab             	&	-19.93	&	3.00	&	-7.65	&	10.16	&	17	&	I, V	\\
  ESO206-014     	&	60.50	&	 SABc            	&	-19.72	&	1.59	&	-9.06	&	8.43	&	17	&	I, V	\\
  ESO206-020A    	&	0.10	&	 E               	&	-8.97	&	1.83	$^	{+	0.07	}	_	{-	0.07	}$	&	-8.82	$^	{+	0.07	}	_	{-	0.07	}$	&	7.29	$^	{+	0.06	}	_	{-	0.07	}$	&	12	&	VIII	\\
  ESO215-009     	&	5.25	&	 I               	&	-14.02	&	1.76	$^	{+	0.13	}	_	{-	0.13	}$	&	-8.66	$^	{+	0.13	}	_	{-	0.13	}$	&	7.32	$^	{+	0.11	}	_	{-	0.13	}$	&	23	&	I	\\
  ESO215-039     	&	61.29	&	 SABc            	&	-21.36	&	2.61	$^	{+	0.64	}	_	{-	0.29	}$	&	-7.91	$^	{+	0.64	}	_	{-	0.29	}$	&	10.86	$^	{+	0.81	}	_	{-	0.29	}$	&	2	&	I	\\
  ESO234-013     	&	60.90	&	 Sbc             	&	-20.12	&	3.00	&	-7.65	&	9.90	&	17	&	I, V	\\
  ESO263-014     	&	69.83	&	 SBb             	&	-21.57	&	2.40	$^	{+	0.79	}	_	{-	0.32	}$	&	-8.12	$^	{+	0.79	}	_	{-	0.32	}$	&	10.78	$^	{+	0.92	}	_	{-	0.32	}$	&	2	&	I	\\
  ESO267-029     	&	76.23	&	 SBab            	&	-20.88	&	2.39	$^	{+	1.16	}	_	{-	0.35	}$	&	-8.14	$^	{+	1.16	}	_	{-	0.35	}$	&	11.12	$^	{+	1.49	}	_	{-	0.35	}$	&	2	&	I	\\
  ESO268-037     	&	68.50	&	 Sc              	&	--	&	2.59	$^	{+	0.11	}	_	{-	0.15	}$	&	-7.94	$^	{+	0.11	}	_	{-	0.15	}$	&	10.95	$^	{+	0.18	}	_	{-	0.15	}$	&	2	&	I	\\
  ESO268-044     	&	49.95	&	 Sb              	&	-19.96	&	3.02	$^	{+	0.57	}	_	{-	0.31	}$	&	-7.50	$^	{+	0.57	}	_	{-	0.31	}$	&	10.62	$^	{+	0.86	}	_	{-	0.31	}$	&	2	&	I	\\
  ESO317-041     	&	81.17	&	 SBbc            	&	-21.09	&	4.41	$^	{+	0.23	}	_	{-	0.27	}$	&	-6.11	$^	{+	0.23	}	_	{-	0.27	}$	&	10.09	$^	{+	0.23	}	_	{-	0.27	}$	&	2	&	I	\\
  ESO322-042     	&	55.99	&	 Sc              	&	-20.66	&	2.13	$^	{+	0.09	}	_	{-	0.09	}$	&	-8.39	$^	{+	0.09	}	_	{-	0.09	}$	&	10.80	$^	{+	0.13	}	_	{-	0.09	}$	&	2	&	I	\\
  ESO323-025     	&	59.76	&	 Sc              	&	--	&	3.13	$^	{+	0.10	}	_	{-	0.08	}$	&	-7.39	$^	{+	0.10	}	_	{-	0.08	}$	&	11.20	$^	{+	0.13	}	_	{-	0.08	}$	&	2	&	I	\\
  ESO323-027     	&	54.90	&	 Sc              	&	-20.99	&	2.69	$^	{+	1.36	}	_	{-	0.30	}$	&	-7.84	$^	{+	1.36	}	_	{-	0.30	}$	&	11.16	$^	{+	1.37	}	_	{-	0.30	}$	&	2	&	I	\\
  ESO323-039     	&	69.90	&	 Sbc             	&	-19.81	&	2.12	$^	{+	0.20	}	_	{-	0.18	}$	&	-8.40	$^	{+	0.20	}	_	{-	0.18	}$	&	10.50	$^	{+	0.29	}	_	{-	0.18	}$	&	2	&	I	\\
  ESO323-042     	&	59.73	&	 Sbc             	&	-21.08	&	2.22	$^	{+	0.38	}	_	{-	0.24	}$	&	-8.31	$^	{+	0.38	}	_	{-	0.24	}$	&	10.75	$^	{+	0.44	}	_	{-	0.24	}$	&	2	&	I	\\
  ESO356-004     	&	0.14	&	 E               	&	-11.76	&	1.68	&	-8.89	&	7.33	&	27	&	IX	\\
  ESO374-003     	&	43.22	&	 SABc            	&	-20.22	&	2.31	$^	{+	0.08	}	_	{-	0.06	}$	&	-8.21	$^	{+	0.08	}	_	{-	0.06	}$	&	10.53	$^	{+	0.11	}	_	{-	0.06	}$	&	2	&	I	\\
  ESO381-005     	&	79.56	&	 SBbc            	&	-20.36	&	2.99	$^	{+	0.11	}	_	{-	0.09	}$	&	-7.54	$^	{+	0.11	}	_	{-	0.09	}$	&	11.22	$^	{+	0.14	}	_	{-	0.09	}$	&	2	&	I	\\
  ESO400-037     	&	37.50	&	 SBc             	&	-19.16	&	3.07	&	-7.58	&	8.94	&	17	&	I, V	\\
  ESO437-004     	&	48.10	&	 Sbc             	&	-20.32	&	2.78	$^	{+	0.13	}	_	{-	0.18	}$	&	-7.74	$^	{+	0.13	}	_	{-	0.18	}$	&	11.03	$^	{+	0.22	}	_	{-	0.18	}$	&	2	&	I	\\
  ESO437-031     	&	56.17	&	 Scd             	&	-19.72	&	2.51	$^	{+	0.08	}	_	{-	0.10	}$	&	-8.01	$^	{+	0.08	}	_	{-	0.10	}$	&	10.68	$^	{+	0.09	}	_	{-	0.10	}$	&	2	&	I	\\
  ESO438-015     	&	49.96	&	 SBbc            	&	-20.75	&	2.22	$^	{+	1.66	}	_	{-	0.29	}$	&	-8.30	$^	{+	1.66	}	_	{-	0.29	}$	&	10.99	$^	{+	1.91	}	_	{-	0.29	}$	&	2	&	I	\\
  ESO439-018     	&	122.20	&	 Sbc             	&	-21.80	&	2.95	$^	{+	0.28	}	_	{-	0.18	}$	&	-7.57	$^	{+	0.28	}	_	{-	0.18	}$	&	11.53	$^	{+	0.34	}	_	{-	0.18	}$	&	2	&	I	\\
  ESO439-020     	&	59.84	&	 SABb            	&	-20.54	&	2.96	$^	{+	0.21	}	_	{-	0.17	}$	&	-7.56	$^	{+	0.21	}	_	{-	0.17	}$	&	11.12	$^	{+	0.27	}	_	{-	0.17	}$	&	2	&	I	\\
  ESO444-047     	&	62.40	&	 SBc             	&	-20.28	&	2.44	$^	{+	0.50	}	_	{-	0.34	}$	&	-8.08	$^	{+	0.50	}	_	{-	0.34	}$	&	10.84	$^	{+	0.81	}	_	{-	0.34	}$	&	2	&	I	\\
  ESO445-015     	&	60.34	&	 Sbc             	&	-20.45	&	2.41	$^	{+	1.48	}	_	{-	0.40	}$	&	-8.12	$^	{+	1.48	}	_	{-	0.40	}$	&	10.78	$^	{+	1.73	}	_	{-	0.40	}$	&	2	&	I	\\
  ESO446-001     	&	98.34	&	 Sbc             	&	-21.42	&	2.09	$^	{+	0.26	}	_	{-	0.16	}$	&	-8.44	$^	{+	0.26	}	_	{-	0.16	}$	&	10.94	$^	{+	0.30	}	_	{-	0.16	}$	&	2	&	I	\\
  ESO501-001     	&	55.57	&	 SABc            	&	--	&	2.31	$^	{+	0.14	}	_	{-	0.23	}$	&	-8.22	$^	{+	0.14	}	_	{-	0.23	}$	&	10.63	$^	{+	0.23	}	_	{-	0.23	}$	&	2	&	I	\\
  IC0342         	&	4.50	&	 SABc            	&	-22.14	&	1.98	&	-8.54	&	10.33	&	1	&	VI	\\
  IC0509         	&	74.00	&	 Sc              	&	-20.79	&	1.28	&	-9.35	&	9.58	&	18	&	I	\\
  IC2233         	&	10.50	&	 SBc             	&	-19.37	&	2.08	$^	{+	0.07	}	_	{-	0.05	}$	&	-8.57	$^	{+	0.07	}	_	{-	0.05	}$	&	9.54	$^	{+	0.15	}	_	{-	0.05	}$	&	7	&	I	\\
  IC2574         	&	4.00	&	 SABm            	&	-18.01	&	1.64	$^	{+	0.68	}	_	{-	0.68	}$	&	-8.94	$^	{+	0.68	}	_	{-	0.68	}$	&	9.66	$^	{+	0.00	}	_	{-	0.68	}$	&	21;7	&	I, III	\\
  IC4298         	&	92.86	&	 Sc              	&	-21.48	&	2.83	$^	{+	0.09	}	_	{-	0.09	}$	&	-7.69	$^	{+	0.09	}	_	{-	0.09	}$	&	11.48	$^	{+	0.11	}	_	{-	0.09	}$	&	2	&	I	\\
  IC4319         	&	66.05	&	 Sbc             	&	-20.77	&	2.72	$^	{+	0.14	}	_	{-	0.13	}$	&	-7.81	$^	{+	0.14	}	_	{-	0.13	}$	&	11.22	$^	{+	0.18	}	_	{-	0.13	}$	&	2	&	I	\\
  IC5152         	&	2.07	&	 IAB             	&	-16.57	&	1.75	$^	{+	0.09	}	_	{-	0.09	}$	&	-8.67	$^	{+	0.09	}	_	{-	0.09	}$	&	8.53	$^	{+	0.09	}	_	{-	0.09	}$	&	23	&	I	\\
  NGC0024        	&	6.80	&	 Sc              	&	-17.86	&	2.19	$^	{+	0.27	}	_	{-	0.27	}$	&	-8.33	$^	{+	0.27	}	_	{-	0.27	}$	&	10.00	$^	{+		}	_	{-	0.27	}$	&	4	&	I	\\
  NGC0045        	&	5.90	&	 SABd            	&	-17.81	&	2.01	$^	{+	0.10	}	_	{-	0.10	}$	&	-8.52	$^	{+	0.10	}	_	{-	0.10	}$	&	9.69	$^	{+		}	_	{-	0.10	}$	&	4	&	I	\\
  NGC0224        	&	0.69	&	 Sb              	&	-20.93	&	2.12	&	-8.45	&	10.84	&	1	&	VI	\\
  NGC0247        	&	2.50	&	 SABc            	&	-18.38	&	2.00	&	-8.57	&	9.88	&	1	&	VI	\\
  NGC0253        	&	4.20	&	 SABc            	&	-21.50	&	2.10	&	-8.47	&	10.74	&	1	&	VI	\\
  NGC0300        	&	1.65	&	 Scd             	&	-17.69	&	1.95	&	-8.62	&	9.48	&	1	&	VI	\\
  NGC0488        	&	48.80	&	 Sb              	&	-22.67	&	2.13	&	-8.44	&	11.40	&	1	&	VI	\\
  NGC0598        	&	0.82	&	 Sc              	&	-18.81	&	2.04	$^	{+	0.10	}	_	{-	0.10	}$	&	-8.55	$^	{+	0.10	}	_	{-	0.10	}$	&	10.03	$^	{+	0.07	}	_	{-	0.10	}$	&	5;16;1	&	I, V, VI	\\
  NGC0628        	&	7.30	&	 Sc              	&	-19.96	&	2.64	&	-7.89	&	10.93	&	10	&	I	\\
  NGC0720        	&	25.70	&	 E               	&	-21.00	&	2.80	$^	{+	0.42	}	_	{-	0.34	}$	&	-7.78	$^	{+	0.42	}	_	{-	0.34	}$	&	11.86	$^	{+	0.42	}	_	{-	0.34	}$	&	9	&	VII	\\
  NGC0821        	&	23.40	&	 E               	&	-20.60	&	2.30	&	-8.28	&	10.81	&	25	&	VIII	\\
  NGC0891        	&	15.60	&	 Sb              	&	-21.27	&	1.91	&	-8.66	&	11.06	&	1	&	VI	\\
  NGC0925        	&	9.20	&	 Scd             	&	-20.04	&	1.93	$^	{+	0.00	}	_	{-	0.00	}$	&	-8.67	$^	{+	0.00	}	_	{-	0.00	}$	&	10.49	$^	{+	0.01	}	_	{-	0.00	}$	&	21;7	&	I, III	\\
  NGC0959        	&	7.80	&	 Sd              	&	-17.22	&	1.89	$^	{+	0.06	}	_	{-	0.06	}$	&	-8.54	$^	{+	0.06	}	_	{-	0.06	}$	&	9.86	$^	{+	0.07	}	_	{-	0.06	}$	&	11	&	III	\\
  NGC1087        	&	30.50	&	 SABc            	&	-21.44	&	2.25	&	-8.31	&	10.40	&	1	&	VI	\\
  NGC1156        	&	7.80	&	 IB              	&	-18.55	&	2.28	&	-8.35	&	8.90	&	18	&	I	\\
  NGC1169        	&	34.40	&	 Sb              	&	-21.61	&	2.37	&	-8.25	&	10.28	&	18	&	I	\\
  NGC1407        	&	26.80	&	 E               	&	-21.76	&	2.92	$^	{+	0.27	}	_	{-	0.28	}$	&	-7.66	$^	{+	0.27	}	_	{-	0.28	}$	&	12.23	$^	{+	0.27	}	_	{-	0.28	}$	&	9	&	VII	\\
  NGC1560        	&	3.00	&	 Scd             	&	-17.31	&	1.86	$^	{+	0.14	}	_	{-	0.14	}$	&	-8.56	$^	{+	0.14	}	_	{-	0.14	}$	&	9.41	$^	{+	0.04	}	_	{-	0.14	}$	&	6	&	I	\\
  NGC2336        	&	48.30	&	 Sbc             	&	-22.79	&	1.78	&	-8.79	&	11.11	&	1	&	VI	\\
  NGC2366        	&	3.41	&	 IB              	&	-17.14	&	1.77	$^	{+	0.07	}	_	{-	0.07	}$	&	-8.78	$^	{+	0.07	}	_	{-	0.07	}$	&	8.75	$^	{+	0.11	}	_	{-	0.07	}$	&	6;21;7;23	&	I, III	\\
  NGC2403        	&	3.22	&	 SABc            	&	-19.41	&	2.29	$^	{+	0.12	}	_	{-	0.12	}$	&	-8.27	$^	{+	0.12	}	_	{-	0.12	}$	&	10.43	$^	{+	0.04	}	_	{-	0.12	}$	&	21;7;1	&	I, III, VI	\\
  NGC2532        	&	39.10	&	 SABc            	&	-20.31	&	2.58	$^	{+	0.00	}	_	{-	0.00	}$	&	-8.04	$^	{+	0.00	}	_	{-	0.00	}$	&	10.25	$^	{+	0.04	}	_	{-	0.00	}$	&	7;18	&	I	\\
  NGC2537        	&	6.90	&	 SBm             	&	-17.21	&	2.50	&	-8.12	&	9.26	&	18	&	I	\\
  NGC2552        	&	10.10	&	 SABm            	&	-18.02	&	2.57	$^	{+	0.00	}	_	{-	0.00	}$	&	-7.96	$^	{+	0.00	}	_	{-	0.00	}$	&	10.36	$^	{+	0.28	}	_	{-	0.00	}$	&	6;11;18;22;7	&	I, III	\\
  NGC2649        	&	58.00	&	 SABb            	&	-20.57	&	2.80	&	-7.82	&	10.59	&	18	&	I	\\
  NGC2841        	&	14.15	&	 Sb              	&	-21.22	&	2.78	$^	{+	0.00	}	_	{-	0.00	}$	&	-7.79	$^	{+	0.00	}	_	{-	0.00	}$	&	11.19	$^	{+	0.17	}	_	{-	0.00	}$	&	21;10;7;1	&	I, III, VI	\\
  NGC2974        	&	20.89	&	 E               	&	-19.99	&	3.43	$^	{+	0.00	}	_	{-	0.00	}$	&	-7.00	$^	{+	0.00	}	_	{-	0.00	}$	&	10.82	$^	{+	0.42	}	_	{-	0.00	}$	&	24	&	II, III	\\
  NGC2976        	&	3.60	&	 Sc              	&	-17.77	&	2.57	$^	{+	0.67	}	_	{-	0.37	}$	&	-8.08	$^	{+	0.67	}	_	{-	0.37	}$	&	9.47	$^	{+	1.24	}	_	{-	0.37	}$	&	7	&	I	\\
  NGC2977        	&	44.10	&	 Sb              	&	-20.52	&	2.70	&	-7.92	&	10.51	&	18	&	I	\\
  NGC3026        	&	21.20	&	 IB              	&	-18.95	&	2.58	&	-8.04	&	10.37	&	18	&	I	\\
  NGC3031        	&	3.60	&	 Sab             	&	-20.71	&	2.38	$^	{+	0.04	}	_	{-	0.04	}$	&	-8.23	$^	{+	0.04	}	_	{-	0.04	}$	&	10.59	$^	{+	0.11	}	_	{-	0.04	}$	&	21;7	&	I, III	\\
  NGC3095        	&	40.32	&	 Sc              	&	-21.47	&	2.54	$^	{+	0.25	}	_	{-	0.14	}$	&	-7.98	$^	{+	0.25	}	_	{-	0.14	}$	&	11.12	$^	{+	0.28	}	_	{-	0.14	}$	&	2	&	I	\\
  NGC3104        	&	10.00	&	 IAB             	&	-16.72	&	1.98	$^	{+	0.07	}	_	{-	0.00	}$	&	-8.67	$^	{+	0.07	}	_	{-	0.00	}$	&	9.41	$^	{+	0.10	}	_	{-	0.00	}$	&	7	&	I	\\
  NGC3109        	&	1.70	&	 SBm             	&	-17.46	&	1.61	&	-8.96	&	8.99	&	1	&	VI	\\
  NGC3184        	&	11.10	&	 SABc            	&	-19.91	&	2.40	&	-8.12	&	10.95	&	10	&	I	\\
  NGC3198        	&	13.88	&	 Sc              	&	-20.77	&	2.14	$^	{+	0.00	}	_	{-	0.00	}$	&	-8.40	$^	{+	0.00	}	_	{-	0.00	}$	&	10.61	$^	{+	0.21	}	_	{-	0.00	}$	&	21;7;10;1	&	I, III, VI	\\
  NGC3274        	&	6.60	&	 Scd             	&	-16.54	&	2.52	$^	{+	0.00	}	_	{-	0.00	}$	&	-8.06	$^	{+	0.00	}	_	{-	0.00	}$	&	9.20	$^	{+	0.08	}	_	{-	0.00	}$	&	6;7;18;20	&	I,II	\\
  NGC3319        	&	14.10	&	 SBc             	&	-19.60	&	2.01	&	-8.61	&	10.00	&	18	&	I	\\
  NGC3346        	&	15.20	&	 SBc             	&	-18.59	&	2.73	&	-7.89	&	10.18	&	18	&	I	\\
  NGC3359        	&	22.80	&	 Sc              	&	-21.04	&	1.75	&	-8.81	&	10.17	&	1	&	VI	\\
  NGC3379        	&	10.30	&	 E               	&	-19.95	&	2.37	&	-8.21	&	10.81	&	25	&	VIII	\\
  NGC3453        	&	57.77	&	 SBbc            	&	-20.62	&	3.57	$^	{+	0.16	}	_	{-	0.12	}$	&	-6.95	$^	{+	0.16	}	_	{-	0.12	}$	&	11.15	$^	{+	0.19	}	_	{-	0.12	}$	&	2	&	I	\\
  NGC3463        	&	56.91	&	 Sb              	&	-20.78	&	3.17	$^	{+	0.49	}	_	{-	0.33	}$	&	-7.35	$^	{+	0.49	}	_	{-	0.33	}$	&	11.06	$^	{+	0.75	}	_	{-	0.33	}$	&	2	&	I	\\
  NGC3521        	&	10.70	&	 SABb            	&	-20.93	&	2.51	$^	{+	0.09	}	_	{-	0.09	}$	&	-8.10	$^	{+	0.09	}	_	{-	0.09	}$	&	10.83	$^	{+	0.05	}	_	{-	0.09	}$	&	21;10;7	&	I, III	\\
  NGC3621        	&	6.60	&	 SBcd            	&	-20.04	&	2.05	$^	{+	0.00	}	_	{-	0.00	}$	&	-8.52	$^	{+	0.00	}	_	{-	0.00	}$	&	10.29	$^	{+	0.01	}	_	{-	0.00	}$	&	21;7	&	I, III	\\
  NGC3627        	&	9.30	&	 SABb            	&	-20.76	&	2.63	&	-7.90	&	11.01	&	10	&	I	\\
  NGC3726        	&	14.30	&	 Sc              	&	-20.47	&	1.68	&	-8.95	&	9.84	&	18	&	I	\\
  NGC3893        	&	15.50	&	 SABc            	&	-20.69	&	3.00	&	-7.63	&	10.68	&	18	&	I	\\
  NGC3898        	&	23.10	&	 Sab             	&	-20.91	&	2.63	&	-7.93	&	10.99	&	1	&	VI	\\
  NGC3992        	&	22.60	&	 Sbc             	&	-21.69	&	2.04	&	-8.53	&	11.12	&	1	&	VI	\\
  NGC4062        	&	11.40	&	 SABc            	&	-19.19	&	2.18	&	-8.44	&	10.02	&	18	&	I	\\
  NGC4125        	&	22.20	&	 E               	&	-21.20	&	2.40	$^	{+	0.17	}	_	{-	0.27	}$	&	-8.17	$^	{+	0.17	}	_	{-	0.27	}$	&	11.66	$^	{+	0.16	}	_	{-	0.27	}$	&	9	&	VII	\\
  NGC4242        	&	8.10	&	 Sd              	&	-18.29	&	2.10	$^	{+	0.06	}	_	{-	0.06	}$	&	-8.54	$^	{+	0.06	}	_	{-	0.06	}$	&	9.59	$^	{+	0.05	}	_	{-	0.06	}$	&	18;7	&	I	\\
  NGC4244        	&	5.00	&	 Sc              	&	-18.74	&	1.70	&	-8.87	&	10.08	&	1	&	VI	\\
  NGC4254        	&	16.10	&	 Sc              	&	-20.85	&	2.40	&	-8.12	&	10.82	&	10	&	I	\\
  NGC4258        	&	10.40	&	 SABb            	&	-21.62	&	2.05	&	-8.52	&	11.07	&	1	&	VI	\\
  NGC4261        	&	29.30	&	 E               	&	-21.09	&	2.18	$^	{+	0.45	}	_	{-	0.29	}$	&	-8.40	$^	{+	0.45	}	_	{-	0.29	}$	&	11.54	$^	{+	0.44	}	_	{-	0.29	}$	&	9	&	VII	\\
  NGC4288        	&	8.40	&	 SBcd            	&	-16.57	&	2.40	$^	{+	0.12	}	_	{-	0.12	}$	&	-8.19	$^	{+	0.12	}	_	{-	0.12	}$	&	9.44	$^	{+	0.25	}	_	{-	0.12	}$	&	19;22;7	&	I,II	\\
  NGC4298        	&	16.10	&	 Sc              	&	-19.57	&	1.86	&	-8.67	&	9.75	&	10	&	I	\\
  NGC4302        	&	16.10	&	 Sc              	&	-19.88	&	2.32	&	-8.20	&	10.92	&	10	&	I	\\
  NGC4303        	&	16.10	&	 Sbc             	&	-21.03	&	2.23	&	-8.29	&	10.97	&	10	&	I	\\
  NGC4321        	&	18.05	&	 SABb            	&	-21.44	&	2.43	$^	{+	0.00	}	_	{-	0.00	}$	&	-8.10	$^	{+	0.00	}	_	{-	0.00	}$	&	11.13	$^	{+	0.26	}	_	{-	0.00	}$	&	10;1	&	I	\\
  NGC4374        	&	16.35	&	 E               	&	-21.18	&	2.77	$^	{+	0.64	}	_	{-	0.75	}$	&	-7.81	$^	{+	0.64	}	_	{-	0.75	}$	&	12.01	$^	{+	0.63	}	_	{-	0.75	}$	&	15	&	VIII	\\
  NGC4395        	&	4.28	&	 Sm              	&	-18.27	&	1.97	$^	{+	0.35	}	_	{-	0.35	}$	&	-8.60	$^	{+	0.35	}	_	{-	0.35	}$	&	9.45	$^	{+	0.54	}	_	{-	0.35	}$	&	6;1;22;7;18	&	I, VI	\\
  NGC4402        	&	16.10	&	 Sb              	&	-18.99	&	1.61	&	-8.91	&	9.85	&	10	&	I	\\
  NGC4455        	&	6.80	&	 SBcd            	&	-16.82	&	1.93	$^	{+	0.00	}	_	{-	0.00	}$	&	-8.62	$^	{+	0.00	}	_	{-	0.00	}$	&	8.74	$^	{+	0.25	}	_	{-	0.00	}$	&	19;6;7	&	I,II	\\
  NGC4472        	&	15.10	&	 E               	&	-21.73	&	2.81	$^	{+	0.14	}	_	{-	0.22	}$	&	-7.76	$^	{+	0.14	}	_	{-	0.22	}$	&	12.24	$^	{+	0.14	}	_	{-	0.22	}$	&	9	&	VII	\\
  NGC4494        	&	15.80	&	 E               	&	-20.42	&	2.15	$^	{+	0.18	}	_	{-	0.18	}$	&	-8.43	$^	{+	0.18	}	_	{-	0.18	}$	&	10.88	$^	{+	0.17	}	_	{-	0.18	}$	&	14	&	VIII	\\
  NGC4501        	&	16.10	&	 Sb              	&	-21.47	&	2.81	&	-7.71	&	11.27	&	10	&	I	\\
  NGC4519        	&	16.10	&	 Scd             	&	-18.98	&	2.24	&	-8.29	&	10.10	&	10	&	I	\\
  NGC4535        	&	16.10	&	 Sc              	&	-20.66	&	2.09	&	-8.43	&	10.81	&	10	&	I	\\
  NGC4536        	&	16.10	&	 SABb            	&	-20.72	&	2.17	&	-8.36	&	10.99	&	10	&	I	\\
  NGC4548        	&	16.10	&	 Sb              	&	-20.40	&	1.87	&	-8.66	&	10.47	&	10	&	I	\\
  NGC4567        	&	16.10	&	 Sbc             	&	-19.25	&	2.35	&	-8.17	&	10.37	&	10	&	I	\\
  NGC4568        	&	16.10	&	 Sbc             	&	-20.07	&	2.51	&	-8.01	&	10.80	&	10	&	I	\\
  NGC4569        	&	16.10	&	 SABa            	&	-21.61	&	2.72	&	-7.80	&	11.36	&	10	&	I	\\
  NGC4575        	&	43.51	&	 SBbc            	&	-20.95	&	2.51	$^	{+	0.42	}	_	{-	0.26	}$	&	-8.01	$^	{+	0.42	}	_	{-	0.26	}$	&	10.90	$^	{+	0.53	}	_	{-	0.26	}$	&	2	&	I	\\
  NGC4579        	&	16.10	&	 SABb            	&	-20.91	&	2.67	&	-7.86	&	11.01	&	10	&	I	\\
  NGC4605        	&	5.03	&	 SBc             	&	-18.45	&	2.58	$^	{+	0.09	}	_	{-	0.09	}$	&	-8.03	$^	{+	0.09	}	_	{-	0.09	}$	&	9.96	$^	{+	0.13	}	_	{-	0.09	}$	&	7;1	&	I,VI	\\
  NGC4635        	&	12.80	&	 SABc            	&	-17.48	&	2.68	&	-7.94	&	9.90	&	18	&	I	\\
  NGC4647        	&	16.10	&	 SABc            	&	-19.19	&	1.92	&	-8.60	&	9.94	&	10	&	I	\\
  NGC4649        	&	15.60	&	 E               	&	-21.27	&	3.14	$^	{+	0.14	}	_	{-	0.26	}$	&	-7.44	$^	{+	0.14	}	_	{-	0.26	}$	&	12.22	$^	{+	0.14	}	_	{-	0.26	}$	&	9	&	VII	\\
  NGC4651        	&	20.60	&	 Sc              	&	-20.61	&	2.54	&	-8.08	&	10.81	&	18	&	I	\\
  NGC4654        	&	16.10	&	 Sc              	&	-20.61	&	2.49	&	-8.03	&	10.89	&	10	&	I	\\
  NGC4689        	&	16.10	&	 Sc              	&	-19.77	&	2.35	&	-8.17	&	10.62	&	10	&	I	\\
  NGC4696A       	&	38.19	&	 SBb             	&	--	&	2.89	$^	{+	1.36	}	_	{-	0.43	}$	&	-7.64	$^	{+	1.36	}	_	{-	0.43	}$	&	11.13	$^	{+	2.03	}	_	{-	0.43	}$	&	2	&	I	\\
  NGC4698        	&	20.00	&	 Sab             	&	-20.49	&	2.51	&	-8.06	&	10.78	&	1	&	VI	\\
  NGC4736        	&	5.15	&	 Sab             	&	-20.02	&	3.56	$^	{+	0.00	}	_	{-	0.00	}$	&	-7.01	$^	{+	0.00	}	_	{-	0.00	}$	&	11.28	$^	{+	0.42	}	_	{-	0.00	}$	&	21;10;7;1	&	I, III, VI	\\
  NGC4789A       	&	4.30	&	 I               	&	-14.44	&	1.73	$^	{+	0.00	}	_	{-	0.00	}$	&	-8.85	$^	{+	0.00	}	_	{-	0.00	}$	&	8.10	$^	{+	0.03	}	_	{-	0.00	}$	&	21;7	&	I, III	\\
  NGC5023        	&	4.80	&	 Sc              	&	-17.18	&	2.38	$^	{+	0.09	}	_	{-	0.09	}$	&	-8.27	$^	{+	0.09	}	_	{-	0.09	}$	&	9.78	$^	{+	0.10	}	_	{-	0.09	}$	&	7	&	I	\\
  NGC5033        	&	17.90	&	 Sc              	&	-21.18	&	2.02	&	-8.54	&	11.08	&	1	&	VI	\\
  NGC5055        	&	10.33	&	 Sbc             	&	-21.18	&	2.03	$^	{+	0.44	}	_	{-	0.44	}$	&	-8.52	$^	{+	0.44	}	_	{-	0.44	}$	&	10.74	$^	{+	0.31	}	_	{-	0.44	}$	&	21;7;10;1	&	I, III, VI	\\
  NGC5194        	&	9.60	&	 Sbc             	&	-21.57	&	2.82	$^	{+	0.00	}	_	{-	0.00	}$	&	-7.74	$^	{+	0.00	}	_	{-	0.00	}$	&	10.79	$^	{+	0.02	}	_	{-	0.00	}$	&	10;1	&	I	\\
  NGC5204        	&	4.86	&	 Sm              	&	-17.14	&	2.25	$^	{+	0.00	}	_	{-	0.00	}$	&	-8.32	$^	{+	0.00	}	_	{-	0.00	}$	&	9.54	$^	{+	0.19	}	_	{-	0.00	}$	&	19;18;20;22;7	&	I,II	\\
  NGC5236        	&	6.90	&	 Sc              	&	-21.69	&	2.00	&	-8.56	&	10.48	&	1	&	VI	\\
  NGC5298        	&	61.59	&	 SBb             	&	-20.74	&	2.97	$^	{+	0.40	}	_	{-	0.29	}$	&	-7.56	$^	{+	0.40	}	_	{-	0.29	}$	&	11.64	$^	{+	0.53	}	_	{-	0.29	}$	&	2	&	I	\\
  NGC5383        	&	46.40	&	 Sb              	&	-21.33	&	2.02	&	-8.55	&	10.68	&	1	&	VI	\\
  NGC5408        	&	4.81	&	 IB              	&	-17.01	&	1.47	$^	{+	0.15	}	_	{-	0.15	}$	&	-8.95	$^	{+	0.15	}	_	{-	0.15	}$	&	8.30	$^	{+	0.11	}	_	{-	0.15	}$	&	23	&	I	\\
  NGC5457        	&	7.60	&	 SABc            	&	-21.11	&	1.91	&	-8.66	&	10.98	&	1	&	VI	\\
  NGC5585        	&	5.70	&	 SABc            	&	-17.84	&	2.43	$^	{+	0.00	}	_	{-	0.00	}$	&	-8.21	$^	{+	0.00	}	_	{-	0.00	}$	&	9.68	$^	{+	0.03	}	_	{-	0.00	}$	&	18;7	&	I	\\
  NGC5608        	&	10.20	&	 IB              	&	-16.77	&	1.96	&	-8.67	&	8.95	&	18	&	I	\\
  NGC5622        	&	54.90	&	 Sb              	&	-20.03	&	2.21	&	-8.42	&	9.86	&	18	&	I	\\
  NGC5727        	&	26.40	&	 SABd            	&	-18.17	&	2.24	$^	{+	0.12	}	_	{-	0.12	}$	&	-8.40	$^	{+	0.12	}	_	{-	0.12	}$	&	9.03	$^	{+	0.02	}	_	{-	0.12	}$	&	18;7	&	I	\\
  NGC5907        	&	15.60	&	 SABc            	&	-21.27	&	2.03	&	-8.53	&	11.09	&	1	&	VI	\\
  NGC5949        	&	10.70	&	 Sbc             	&	-17.95	&	2.55	$^	{+	0.16	}	_	{-	0.16	}$	&	-8.09	$^	{+	0.16	}	_	{-	0.16	}$	&	9.59	$^	{+	0.03	}	_	{-	0.16	}$	&	7;18	&	I	\\
  NGC5963        	&	15.00	&	 Sbc             	&	-18.07	&	2.65	$^	{+	0.00	}	_	{-	0.00	}$	&	-7.98	$^	{+	0.00	}	_	{-	0.00	}$	&	10.28	$^	{+	0.04	}	_	{-	0.00	}$	&	7;1	&	I,VI	\\
  NGC6015        	&	14.70	&	 Sc              	&	-19.91	&	2.86	$^	{+	0.00	}	_	{-	0.00	}$	&	-7.78	$^	{+	0.00	}	_	{-	0.00	}$	&	10.81	$^	{+	0.02	}	_	{-	0.00	}$	&	18;7	&	I	\\
  NGC6482        	&	58.80	&	 E               	&	-22.11	&	2.79	$^	{+	0.40	}	_	{-	0.18	}$	&	-7.79	$^	{+	0.40	}	_	{-	0.18	}$	&	11.89	$^	{+	0.39	}	_	{-	0.18	}$	&	9	&	VII	\\
  NGC6503        	&	6.10	&	 Sc              	&	-19.03	&	2.14	&	-8.43	&	9.79	&	1	&	VI	\\
  NGC6689        	&	11.00	&	 SABc            	&	-18.42	&	2.45	$^	{+	0.05	}	_	{-	0.07	}$	&	-8.20	$^	{+	0.05	}	_	{-	0.07	}$	&	10.05	$^	{+	0.08	}	_	{-	0.07	}$	&	7	&	I	\\
  NGC6822        	&	0.49	&	 IB              	&	-15.16	&	1.82	$^	{+	0.02	}	_	{-	0.02	}$	&	-8.61	$^	{+	0.02	}	_	{-	0.02	}$	&	8.26	$^	{+	0.02	}	_	{-	0.02	}$	&	26	&	III	\\
  NGC6946        	&	6.10	&	 SABc            	&	-20.70	&	2.25	$^	{+	0.02	}	_	{-	0.02	}$	&	-8.32	$^	{+	0.02	}	_	{-	0.02	}$	&	10.49	$^	{+	0.12	}	_	{-	0.02	}$	&	21;10;7;1	&	I, III, VI	\\
  NGC7137        	&	22.50	&	 SABc            	&	-19.39	&	2.33	$^	{+	0.82	}	_	{-	0.82	}$	&	-8.09	$^	{+	0.82	}	_	{-	0.82	}$	&	10.03	$^	{+	0.49	}	_	{-	0.82	}$	&	11	&	III	\\
  NGC7217        	&	19.75	&	 Sab             	&	-20.94	&	2.86	$^	{+	0.00	}	_	{-	0.00	}$	&	-7.74	$^	{+	0.00	}	_	{-	0.00	}$	&	11.13	$^	{+	0.28	}	_	{-	0.00	}$	&	1;18	&	VI,I	\\
  NGC7331        	&	17.23	&	 Sbc             	&	-22.00	&	2.34	$^	{+	0.00	}	_	{-	0.00	}$	&	-8.22	$^	{+	0.00	}	_	{-	0.00	}$	&	11.16	$^	{+	0.06	}	_	{-	0.00	}$	&	7;10;1	&	I,VI	\\
  NGC7541        	&	57.50	&	 SBc             	&	-22.41	&	2.43	&	-8.13	&	11.35	&	1	&	VI	\\
  NGC7589        	&	123.00	&	 SABa            	&	-20.75	&	1.75	$^	{+	0.05	}	_	{-	0.05	}$	&	-8.83	$^	{+	0.05	}	_	{-	0.05	}$	&	11.60	$^	{+	0.06	}	_	{-	0.05	}$	&	8	&	III	\\
  NGC7664        	&	74.20	&	 Sc              	&	-21.60	&	2.21	&	-8.36	&	10.88	&	1	&	VI	\\
  NGC7793        	&	4.00	&	 Scd             	&	-18.83	&	2.34	$^	{+	0.01	}	_	{-	0.01	}$	&	-8.24	$^	{+	0.01	}	_	{-	0.01	}$	&	10.00	$^	{+	0.14	}	_	{-	0.01	}$	&	21;7;1	&	I, III, VI	\\
  PGC042102      	&	322.00	&	                 	&	-19.68	&	2.39	$^	{+	0.25	}	_	{-	0.25	}$	&	-8.19	$^	{+	0.25	}	_	{-	0.25	}$	&	12.20	$^	{+	0.34	}	_	{-	0.25	}$	&	8	&	III	\\
  PGC044532      	&	7.50	&	 Sm              	&	-17.34	&	1.34	$^	{+	0.14	}	_	{-	0.14	}$	&	-9.09	$^	{+	0.14	}	_	{-	0.14	}$	&	8.96	$^	{+	0.12	}	_	{-	0.14	}$	&	23	&	I	\\
  PGC086613      	&	45.00	&	 Sd              	&	-16.04	&	2.10	$^	{+	0.00	}	_	{-	0.00	}$	&	-8.32	$^	{+	0.00	}	_	{-	0.00	}$	&	10.42	$^	{+	0.00	}	_	{-	0.00	}$	&	6;13	&	I, X	\\
  PGC086619      	&	85.00	&	 Sc              	&	-17.82	&	2.13	&	-8.29	&	10.83	&	13	&	X	\\
  PGC086620      	&	77.00	&	 Sd              	&	-18.50	&	1.92	$^	{+	0.00	}	_	{-	0.00	}$	&	-8.50	$^	{+	0.00	}	_	{-	0.00	}$	&	10.66	$^	{+	0.05	}	_	{-	0.00	}$	&	13;20	&	X, II	\\
  PGC086622      	&	201.00	&	 Scd             	&	-21.55	&	1.98	&	-8.60	&	12.25	&	28	&	IV	\\
  PGC086633      	&	48.00	&	 Sc              	&	-17.94	&	2.21	&	-8.21	&	11.21	&	13	&	X	\\
  PGC086638      	&	96.00	&	 Sd              	&	--	&	1.48	$^	{+	0.00	}	_	{-	0.00	}$	&	-8.94	$^	{+	0.00	}	_	{-	0.00	}$	&	10.35	$^	{+	0.41	}	_	{-	0.00	}$	&	20;13	&	II, X	\\
  PGC086657      	&	32.00	&	 Sm              	&	-16.44	&	1.98	$^	{+	0.63	}	_	{-	0.63	}$	&	-8.44	$^	{+	0.63	}	_	{-	0.63	}$	&	10.20	$^	{+	0.26	}	_	{-	0.63	}$	&	13;11	&	III, X	\\
  PGC086660      	&	49.00	&	 Sc              	&	--	&	1.56	&	-8.86	&	10.03	&	13	&	X	\\
  PGC086666      	&	61.00	&	 I               	&	-17.25	&	2.04	&	-8.38	&	10.22	&	13	&	X	\\
  PGC086670      	&	48.00	&	 I               	&	--	&	1.34	&	-9.09	&	9.72	&	13	&	X	\\
  PGC086672      	&	80.00	&	 Sd              	&	-16.37	&	1.85	$^	{+	0.00	}	_	{-	0.00	}$	&	-8.58	$^	{+	0.00	}	_	{-	0.00	}$	&	10.44	$^	{+	0.12	}	_	{-	0.00	}$	&	20;13	&	II, X	\\
  PGC086674      	&	79.00	&	 Sm              	&	--	&	1.60	&	-8.82	&	10.09	&	13	&	X	\\
  PGC088608      	&	0.09	&	 E-S0            	&	-7.86	&	2.36	$^	{+	0.19	}	_	{-	0.19	}$	&	-8.29	$^	{+	0.19	}	_	{-	0.19	}$	&	7.65	$^	{+	0.22	}	_	{-	0.19	}$	&	12	&	VIII	\\
  PGC166192      	&	5.60	&	 I               	&	-14.03	&	1.88	$^	{+	0.17	}	_	{-	0.17	}$	&	-8.55	$^	{+	0.17	}	_	{-	0.17	}$	&	8.13	$^	{+	0.04	}	_	{-	0.17	}$	&	3	&	III	\\
  UGC00128       	&	60.00	&	 Sd              	&	--	&	1.74	$^	{+	0.00	}	_	{-	0.00	}$	&	-8.69	$^	{+	0.00	}	_	{-	0.00	}$	&	10.80	$^	{+	0.17	}	_	{-	0.00	}$	&	13;11	&	III, X	\\
  UGC00191       	&	17.60	&	 Sm              	&	-17.87	&	2.31	$^	{+	0.07	}	_	{-	0.07	}$	&	-8.12	$^	{+	0.07	}	_	{-	0.07	}$	&	9.97	$^	{+	0.10	}	_	{-	0.07	}$	&	11	&	III	\\
  UGC00731       	&	8.00	&	 I               	&	-15.18	&	2.21	$^	{+	0.00	}	_	{-	0.00	}$	&	-8.40	$^	{+	0.00	}	_	{-	0.00	}$	&	9.35	$^	{+	0.14	}	_	{-	0.00	}$	&	6;22;7	&	I	\\
  UGC01230       	&	51.00	&	 Sm              	&	--	&	2.08	$^	{+	0.00	}	_	{-	0.00	}$	&	-8.50	$^	{+	0.00	}	_	{-	0.00	}$	&	10.47	$^	{+	0.06	}	_	{-	0.00	}$	&	13;6;7	&	X,I	\\
  UGC01281       	&	5.50	&	 Sd              	&	-17.23	&	1.94	$^	{+	0.09	}	_	{-	0.07	}$	&	-8.71	$^	{+	0.09	}	_	{-	0.07	}$	&	9.38	$^	{+	0.16	}	_	{-	0.07	}$	&	7	&	I	\\
  UGC01551       	&	20.20	&	 SBc             	&	-18.49	&	1.94	$^	{+	0.13	}	_	{-	0.13	}$	&	-8.49	$^	{+	0.13	}	_	{-	0.13	}$	&	10.01	$^	{+	0.19	}	_	{-	0.13	}$	&	11	&	III	\\
  UGC02034       	&	10.10	&	 IAB             	&	-16.73	&	2.36	&	-8.26	&	8.56	&	18	&	I	\\
  UGC03137       	&	18.40	&	 Sbc             	&	-17.72	&	1.98	$^	{+	0.05	}	_	{-	0.05	}$	&	-8.67	$^	{+	0.05	}	_	{-	0.05	}$	&	10.20	$^	{+	0.06	}	_	{-	0.05	}$	&	7	&	I	\\
  UGC03371       	&	12.80	&	 I               	&	-17.03	&	1.87	$^	{+	0.10	}	_	{-	0.10	}$	&	-8.70	$^	{+	0.10	}	_	{-	0.10	}$	&	10.15	$^	{+	0.12	}	_	{-	0.10	}$	&	6;22;7	&	I	\\
  UGC03876       	&	14.50	&	 Scd             	&	-17.43	&	3.46	&	-7.16	&	9.79	&	18	&	I	\\
  UGC03974       	&	4.00	&	 IB              	&	-14.46	&	2.27	$^	{+	0.52	}	_	{-	0.52	}$	&	-8.15	$^	{+	0.52	}	_	{-	0.52	}$	&	9.69	$^	{+	0.17	}	_	{-	0.52	}$	&	6	&	I	\\
  UGC04173       	&	16.80	&	 I               	&	-15.47	&	1.25	$^	{+	--	}	_	{-	--	}$	&	-9.31	$^	{+	--	}	_	{-	--	}$	&	9.91	$^	{+	0.07	}	_	{-	--	}$	&	6;7	&	I	\\
  UGC04305       	&	3.40	&	 I               	&	-16.92	&	1.69	$^	{+	0.25	}	_	{-	0.25	}$	&	-8.96	$^	{+	0.25	}	_	{-	0.25	}$	&	8.82	$^	{+	0.29	}	_	{-	0.25	}$	&	7	&	I	\\
  UGC04459       	&	3.60	&	 I               	&	-13.48	&	1.66	$^	{+	0.55	}	_	{-	0.32	}$	&	-8.99	$^	{+	0.55	}	_	{-	0.32	}$	&	6.46	$^	{+	3.47	}	_	{-	0.32	}$	&	7	&	I	\\
  UGC04499       	&	13.00	&	 Sd              	&	-15.83	&	1.86	$^	{+	1.76	}	_	{-	1.76	}$	&	-8.70	$^	{+	1.76	}	_	{-	1.76	}$	&	9.42	$^	{+	0.18	}	_	{-	1.76	}$	&	19;22;20;7;18	&	I,II	\\
  UGC05005       	&	52.00	&	 I               	&	-17.69	&	1.76	$^	{+	0.25	}	_	{-	0.25	}$	&	-8.66	$^	{+	0.25	}	_	{-	0.25	}$	&	10.42	$^	{+	0.10	}	_	{-	0.25	}$	&	6;13	&	I, X	\\
  UGC05139       	&	3.80	&	 IAB             	&	-14.78	&	1.99	$^	{+	0.07	}	_	{-	0.07	}$	&	-8.66	$^	{+	0.07	}	_	{-	0.07	}$	&	8.43	$^	{+	0.10	}	_	{-	0.07	}$	&	7	&	I	\\
  UGC05272       	&	6.43	&	 IB              	&	-15.84	&	2.45	$^	{+	0.00	}	_	{-	0.00	}$	&	-8.11	$^	{+	0.00	}	_	{-	0.00	}$	&	9.91	$^	{+	0.44	}	_	{-	0.00	}$	&	11;6;18	&	III,I	\\
  UGC05423       	&	5.30	&	 I               	&	-14.73	&	2.32	$^	{+	0.07	}	_	{-	0.09	}$	&	-8.33	$^	{+	0.07	}	_	{-	0.09	}$	&	8.30	$^	{+	0.10	}	_	{-	0.09	}$	&	7	&	I	\\
  UGC05470       	&	0.24	&	 E               	&	-10.90	&	2.16	$^	{+	0.11	}	_	{-	0.11	}$	&	-8.49	$^	{+	0.11	}	_	{-	0.11	}$	&	7.54	$^	{+	0.11	}	_	{-	0.11	}$	&	12	&	VIII	\\
  UGC05750       	&	56.00	&	 SBd             	&	-17.74	&	1.46	$^	{+	0.90	}	_	{-	0.90	}$	&	-8.96	$^	{+	0.90	}	_	{-	0.90	}$	&	10.43	$^	{+	0.04	}	_	{-	0.90	}$	&	13;6	&	X, I	\\
  UGC05999       	&	45.00	&	 I               	&	-17.88	&	2.25	&	-8.18	&	10.98	&	13	&	X	\\
  UGC06253       	&	0.23	&	 E               	&	-9.28	&	2.04	$^	{+	0.14	}	_	{-	0.14	}$	&	-8.61	$^	{+	0.14	}	_	{-	0.14	}$	&	7.42	$^	{+	0.14	}	_	{-	0.14	}$	&	12	&	VIII	\\
  UGC06446       	&	12.00	&	 Scd             	&	-17.11	&	2.00	$^	{+	0.26	}	_	{-	0.26	}$	&	-8.60	$^	{+	0.26	}	_	{-	0.26	}$	&	9.15	$^	{+	0.24	}	_	{-	0.26	}$	&	19;22;7	&	I,II	\\
  UGC07559       	&	3.20	&	 I               	&	-13.77	&	1.72	$^	{+	0.00	}	_	{-	0.00	}$	&	-8.91	$^	{+	0.00	}	_	{-	0.00	}$	&	8.55	$^	{+	0.03	}	_	{-	0.00	}$	&	22;7	&	I	\\
  UGC07699       	&	9.30	&	 SBc             	&	-17.53	&	1.99	&	-8.63	&	9.54	&	18	&	I	\\
  UGC09211       	&	12.60	&	 I               	&	-15.18	&	1.95	$^	{+	0.34	}	_	{-	0.34	}$	&	-8.62	$^	{+	0.34	}	_	{-	0.34	}$	&	8.98	$^	{+	0.39	}	_	{-	0.34	}$	&	19;6;22;7	&	I,II	\\
  UGC09749       	&	0.07	&	 E               	&	-3.95	&	2.41	$^	{+	0.00	}	_	{-	0.00	}$	&	-8.20	$^	{+	0.00	}	_	{-	0.00	}$	&	7.62	$^	{+	0.19	}	_	{-	0.00	}$	&	12;27	&	VIII, IX	\\
  UGC10310       	&	14.15	&	 Sm              	&	-17.31	&	1.85	$^	{+	0.14	}	_	{-	0.14	}$	&	-8.79	$^	{+	0.14	}	_	{-	0.14	}$	&	9.49	$^	{+	0.08	}	_	{-	0.14	}$	&	18;7	&	I	\\
  UGC10822       	&	0.08	&	 E               	&	-7.90	&	2.05	$^	{+	0.00	}	_	{-	0.00	}$	&	-8.58	$^	{+	0.00	}	_	{-	0.00	}$	&	7.71	$^	{+	0.09	}	_	{-	0.00	}$	&	12;27	&	VIII, IX	\\
  UGC11557       	&	19.70	&	 SABd            	&	-18.26	&	2.08	$^	{+	0.20	}	_	{-	0.20	}$	&	-8.55	$^	{+	0.20	}	_	{-	0.20	}$	&	9.81	$^	{+	0.01	}	_	{-	0.20	}$	&	18;7	&	I	\\
  UGC11583       	&	5.60	&	 I               	&	-15.35	&	1.74	$^	{+	0.03	}	_	{-	0.03	}$	&	-8.68	$^	{+	0.03	}	_	{-	0.03	}$	&	8.54	$^	{+	0.03	}	_	{-	0.03	}$	&	3	&	III	\\
  UGC11707       	&	15.90	&	 Sd              	&	-16.52	&	2.13	$^	{+	0.00	}	_	{-	0.00	}$	&	-8.48	$^	{+	0.00	}	_	{-	0.00	}$	&	9.78	$^	{+	0.44	}	_	{-	0.00	}$	&	20;22;7;18	&	I, II	\\
  UGC11820       	&	13.30	&	 SABm            	&	--	&	1.93	$^	{+	0.11	}	_	{-	0.11	}$	&	-8.49	$^	{+	0.11	}	_	{-	0.11	}$	&	10.65	$^	{+	0.15	}	_	{-	0.11	}$	&	11	&	III	\\
  UGC11861       	&	25.10	&	 SABd            	&	-20.08	&	2.05	$^	{+	0.00	}	_	{-	0.00	}$	&	-8.53	$^	{+	0.00	}	_	{-	0.00	}$	&	10.33	$^	{+	0.34	}	_	{-	0.00	}$	&	20;22	&	I, II	\\
  UGC12060       	&	15.70	&	 IB              	&	-16.37	&	2.41	$^	{+	0.26	}	_	{-	0.26	}$	&	-8.21	$^	{+	0.26	}	_	{-	0.26	}$	&	9.49	$^	{+	0.24	}	_	{-	0.26	}$	&	18;22;7	&	I	\\
  UGC12632       	&	6.90	&	 SABm            	&	-17.12	&	2.18	$^	{+	0.00	}	_	{-	0.00	}$	&	-8.45	$^	{+	0.00	}	_	{-	0.00	}$	&	9.75	$^	{+	0.13	}	_	{-	0.00	}$	&	22;7	&	I	\\
  UGC12732       	&	13.20	&	 SABm            	&	-16.77	&	1.87	$^	{+	0.00	}	_	{-	0.00	}$	&	-8.70	$^	{+	0.00	}	_	{-	0.00	}$	&	9.72	$^	{+	0.24	}	_	{-	0.00	}$	&	19;20;22;7	&	I,II	\\

\end{longtable}
\end{document}